\documentclass[twocolumn]{aastex62}
\pdfoutput=1

\usepackage{amsmath,amssymb}

\newcommand{\new}[1]{{#1}}

\DeclareMathOperator\erf{erf}

\newcommand{\intd}{ \, \mathrm{d} }

\newcommand{\water}{ \mathrm{H_{2}O} }
\newcommand{\sil}{ \mathrm{SiO_2} }
\newcommand{\co}{ \mathrm{CO} }
\newcommand{\coo}{ \mathrm{CO_2} }

\newcommand{\uv}{ \mathrm{UV} }

\newcommand{\sat}{ \mathrm{sat} }

\newcommand{\hy}{ \mathrm{H} }
\newcommand{\hyy}{ \mathrm{H_{2}} }
\newcommand{\ice}{ \mathrm{ice}}
\newcommand{\St}{ \mathrm{St}}
\newcommand{\Sc}{ \mathrm{Sc}}
\newcommand{\RET}{ \mathrm{Re_T}}

\newcommand{\Cv}{ C_\mathrm{v}}
\newcommand{\Cd}{ C_\mathrm{d}}
\newcommand{\Cice}{ C_\mathrm{ice}}

\newcommand{\fw}{f_\mathrm{w}}

\begin{document}

\title{Transport of CO in Protoplanetary Disks:\\ Consequences of Pebble Formation, Settling, and Radial Drift}

\author{Sebastiaan Krijt}
\altaffiliation{Hubble Fellow}
\affiliation{Department of the Geophysical Sciences, The University of Chicago, 5734 South Ellis Avenue, Chicago, IL 60637, USA}

\author{Kamber R. Schwarz}
\affiliation{Department of Astronomy, University of Michigan, 500 Church Street, Ann Arbor, Michigan 48109, USA}

\author{Edwin A. Bergin}
\affiliation{Department of Astronomy, University of Michigan, 500 Church Street, Ann Arbor, Michigan 48109, USA}

\author{Fred J. Ciesla}
\affiliation{Department of the Geophysical Sciences, The University of Chicago, 5734 South Ellis Avenue, Chicago, IL 60637, USA}

\correspondingauthor{Sebastiaan Krijt}
\email{skrijt@uchicago.edu}

\begin{abstract} 
Current models of (exo)planet formation often rely on a large influx of so-called `pebbles' from the outer disk into the planet formation region. In this paper, we investigate how the formation of pebbles in the cold outer regions of protoplanetary disks and their subsequent migration to the inner disk can alter the gas-phase CO distribution both interior and exterior to the midplane CO snowline. By simulating the resulting CO abundances in the midplane as well as the warm surface layer, we identify observable signatures of large-scale pebble formation and migration that can be used as `smoking guns' for these important processes. Specifically, we find that after $1\mathrm{~Myr}$, the formation and settling of icy pebbles results in the removal of up to $80\%$ of the CO vapor in the warm ($T>22\mathrm{~K}$) disk layers outside the CO snowline, while the radial migration of pebbles results in the generation of a plume of CO vapor interior the snowline, increasing the CO abundance by a factor ${\sim}2{-}6$ depending on the strength of the turbulence and the sizes of the individual pebbles. The absence of this plume of CO vapor in young nearby disks could indicate efficient conversion of CO into a more refractory species, or a reduction in the radial mass flux of pebbles by, for example, disk inhomogeneities or early planetesimal formation.
\end{abstract}

\keywords{protoplanetary disks --- astrochemistry --- stars: circumstellar matter --- methods: numerical}

\section{Introduction}
Snowlines are believed to play an important role in protoplanetary disk evolution and planet formation in general. Marking the locations where major volatiles (e.g., $\water,\co,\coo$) transition from being predominantly in the gas-phase to solid as ices on grain surfaces, snowlines separate regions of the protoplanetary disk with possibly very different gas-phase and grain-surface chemistry, changes that are often assumed to be reflected in the composition of (giant) planets forming in different locations \citep[e.g.,][]{oberg2011}. 

The formation of planetesimals and planetary embryos is often associated with the water snowline \citep[e.g.,][]{drazkowska2017,schoonenberg2017,ormel2017}, but other snowlines could also be preferred sites \citep{ali-dib2017}. In the popular `pebble accretion' paradigm, planetesimals/embryos then grow rapidly by accreting mm/cm-size pebbles that drift in from further out in the disk \citep{ormelklahr2010,lambrechts2012,johansen2017}. While growth through pebble accretion can be very fast, only a small fraction of pebbles is usually accreted \citep{ormel2018}, and therefore the process relies on a large and long-lived radial flux of pebbles coming in from the outer regions of the protoplanetary nebula \citep{lambrechts2014}. Such a large-scale radial migration of ice-covered solids originating from the outer disk is expected to redistribute volatiles on a disk-wide scale \citep{oberg2016}, qualitatively changing the static picture presented in \citet{oberg2011}.

The interaction between midplane snowlines and radial transport of solids and vapor has been studied in the past \citep{stevenson1988,cuzzi2004,ciesla2006} and has received a lot of attention in recent years \citep{stammler2017,schoonenberg2017,booth2017,drazkowska2017,bosman2017}. With radial drift being faster than turbulent mixing, these studies generally find an enhancement of volatiles interior to their snowline, the magnitude of which depends on the underlying pebble flux and ice content.

Even before pebbles start drifting however, the formation of these large, settled dust particles can change the vertical distribution of gas-phase volatiles via the sequestration of ices in the midplane \citep{meijerink2009,du2015,kama2016,du2017}. Models studying vertical mixing find that this effect can decrease the gas-phase $\water$ and CO abundances in the warm molecular layer by anywhere between a factor of a few to almost 2 orders of magnitude, depending on the timescales involved and the details of the pebble formation process \citep{xu2016,krijt2016b}.

For CO, this story of depletion above the surface snowline and potential enhancement in the inner disk is of particular importance because CO emission is commonly used as a tracer for bulk disk mass \citep[e.g.,][]{williams2014,ansdell2016,miotello2016,miotello2017,molyarova2017}. Hence, if the CO abundance is significantly depleted in the region of the disk that dominates the emission, this approach could be underestimating the true disk mass. For the handful of disks for which independent mass estimates can be made using HD, it appears as though CO is indeed depleted by a factor of a few to up to two orders of magnitude \citep{favre2013,mcclure2016,schwarz2016}. In addition, CO is the only molecule for which the snowline has been (directly) observed \citep{qi2013} and for which we can vertically and radially resolve abundances using a variety of isotopologues \citep{schwarz2016,zhang2017,dutrey2017,pinte2017,huang2018}.

The aim of this paper is to construct a self-consistent model that describes how the formation and subsequent vertical settling and radial drift of pebbles alters CO abundances in different regions of the disk; both interior and exterior to the midplane snowline, as well as in the warmer surface layers of the outer disk. To that end, we focus on a single, invariant disk profile (Sect.~\ref{sec:model}) and model the vertical and radial transport of dust, pebbles, ices, and gas-phase CO while pebbles are continuously forming over Myr timescales (Sect.~\ref{sec:numerical}). By comparing models of increasing complexity (Sect.~\ref{sec:building}) and exploring the dependence on several parameters related to pebble formation/evolution (Sect.~\ref{sec:results}), we attempt to build a coherent story of how pebble migration affects CO abundances on a disk-wide scale. The results are discussed in Sect. \ref{sec:discussion} and conclusions presented in Sect. \ref{sec:concl}.

\section{Model}\label{sec:model}
Here we describe the physical and thermal structure of the disk (Sect.~\ref{sec:diskmodel}), the equations governing transport of gas-phase molecules and solids (Sect.~\ref{sec:transport}), and conditions and rates at which freeze-out and desorption of CO in different environments (Sect.~\ref{sec:FOandDES}). Finally, in Sect.~\ref{sec:localpebbleformation}, we outline how small dust grains coagulate to form pebbles.

\subsection{Disk structure}\label{sec:diskmodel}
We focus on a disk around a $1M_\odot$ star, with a radial gas surface density profile \citep{lyndenbell1974,hartmann1998}
\begin{equation}\label{eq:sigma_g}
\Sigma_\mathrm{g}(r) = \Sigma_c \left( \frac{r}{r_c}\right)^{-p} \exp\left\{ -\left(\frac{r}{r_c}\right)^{2-p} \right\},
\end{equation}
which is normalized by choosing a total disk mass
\begin{equation}
\Sigma_c = (2-p) \frac{M_\mathrm{disk}}{2\pi r_c^2}.
\end{equation}
For $p=1$, such a profile contains 10\%, 63\%, 86\%, and 95\% of the disk's mass within 0.1, 1, 2, and $3r_c$, respectively. For the midplane temperature, we assume
\begin{equation}\label{eq:T_mid}
T_\mathrm{mid}(r) = T_0 \left( \frac{r}{\mathrm{au}} \right)^{-1/2},
\end{equation}
with $T_0=130 \mathrm{~K}$. The vertical density structure is then parametrized as follows
\begin{equation}\label{eq:rho_g}
\rho_\mathrm{g}(r,z) = \frac{\Sigma_\mathrm{g}(r)}{\sqrt{2\pi} H} \exp \left\{ - \frac{1}{2} \left( \frac{z}{H} \right)^2\right\}
\end{equation}
where the scale-height is given by $H = c_s / \Omega$, the soundspeed $c_s = \sqrt{ k_\mathrm{B} T_\mathrm{mid}(r) / \mu m_\mathrm{H} }$, and $\mu =2.3$ is the mean molecular weight. Temperatures at the disk surface are elevated as they are directly exposed to warming radiation \citep{chiang1997}; here we adopt a vertical temperature sturcture that is based on \citet{rosenfeld2013,dutrey2017}: above $z=z_qH$, the atmospheric temperature is parametrized as
\begin{equation}
T_\mathrm{atm}(r)=3T_\mathrm{mid}(r),
\end{equation}
and for smaller $z$
\begin{equation}
T(r,z) = T_\mathrm{mid}(r) + (T_\mathrm{atm}(r)-T_\mathrm{mid}(r)) \left[\sin \left( \frac{\pi z}{2z_q H}  \right)\right]^{2\delta},
\end{equation}
where we will use $z_q=3$ and $\delta=2$. We also calculate the cumulative UV vertical optical depth. Assuming that dust is the primary absorber of radiations, with an opacity of $\kappa_\uv = 3/(4s_\bullet \rho_\bullet)$, where $s_\bullet$ and $\rho_\bullet$ are the monomer size and material density, the cumulative optical depth can be calculated as
\begin{equation}\label{eq:tau_0}
\begin{split}
\tau_\uv(z) =& \int_z^\infty \kappa_\uv  \rho_\mathrm{d}(z)  \intd z \\ =& \kappa_\uv \frac{\Sigma_\mathrm{g}}{2}  \frac{\rho_\mathrm{d}}{\rho_\mathrm{g}} \left( 1-\erf \left\{ \frac{z/H}{\sqrt{2}}\right\} \right).
\end{split}
\end{equation}
\new{The analytical solution in Eq.~\ref{eq:tau_0} is only valid when the dust-to-gas ratio is constant with height. During our simulations, however, the small grain abundance will vary in time and space and we evaluate the integral in Eq.~\ref{eq:tau_0} numerically.}

\begin{figure}[t]
\centering
\includegraphics[clip=,width=1.\linewidth]{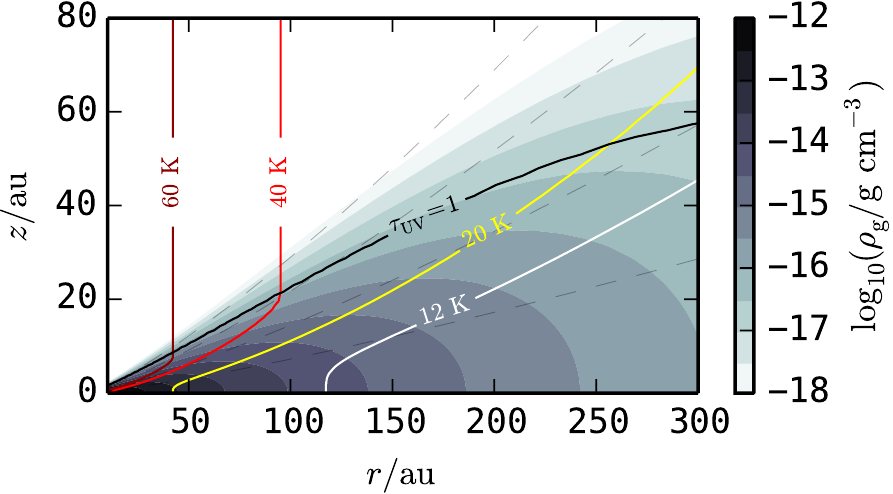}
\caption{Bulk gas density and temperature for our standard disk model (see Sec. \ref{sec:diskmodel}). The black contour denotes the $\tau_\mathrm{UV}=1$ surface (Eq.~\ref{eq:tau_0}, and dashed lines show $z/H=\{1,2,3,4\}$.}
\label{fig:diskmodel}
\end{figure}

\new{In this paper, we focus exclusively on a disk model with $r_c = 100\mathrm{~au}$, $p=1$, and $M_\mathrm{d} = 0.05~M_\odot$. Figure \ref{fig:diskmodel} shows the gas density (Eq.~\ref{eq:rho_g}) and temperature structure for these parameters. Dust is initially present at a dust-to-gas ratio of 1/100, with all grains being $s_\bullet=0.1\mathrm{~\mu m}$ in size with a material density of $\rho_\bullet=2\mathrm{~g/cm^3}$. For these numbers, $\kappa_\mathrm{uv}\approx 8\times10^{4}\mathrm{~cm^2/g}$. The colored contours show various temperatures as well as the $\tau_\mathrm{UV}=1$ surface.}

\subsection{Transport of vapor and solids}\label{sec:transport}
\new{The turbulent viscosity in the gas disk is parametrized as $\nu_\mathrm{T} = \alpha c_s H$ \citep{shakura1973}, with $\alpha$ assumed constant in the radial and vertical direction and $c_s$ evaluated at the midplane. This viscosity influences the transport of material in two ways. First, gas will move towards the star at an accretion rate of $\dot{M}_\mathrm{g} = 3\pi \nu_\mathrm{T} \Sigma_\mathrm{g}$, and a local viscous timescale can be estimated as $t_\nu \sim r^2 / \nu_\mathrm{T}$ \citep[e.g.,][]{hartmann1998}. For the disk profile of Eq.~\ref{eq:sigma_g} in combination with $\alpha=10^{-3}$, we obtain $\dot{M}\sim10^{-9}~M_\odot/\mathrm{yr}$ and $t_\nu > 3\mathrm{~Myr}$ for radii $r>10\mathrm{~au}$. As we will be limiting our simulations to a period of $1\mathrm{~Myr}$, we ignore the effects of the disk's viscous evolution and treat the bulk of the gas as being static for simplicity.}

\new{The second consequence of the presence of a viscosity is that the associated diffusion will act to smear out concentration gradients present in gas-phase and/or dust species. In the case where the gas density does not evolve in time,} transport equations for a trace species with concentration $C_i \equiv \rho_i / \rho_\mathrm{g} \ll 1$ are given by \citep{ciesla2009}
\begin{equation}\label{eq:dCdt_motion}
\begin{split}
 \dfrac{ \partial C_i }{\partial t} = \frac{1}{r \rho_\mathrm{g}} \dfrac{\partial}{\partial r }\left( r \rho_g D_i \dfrac{\partial C_i}{\partial r} \right) -    \frac{1}{r \rho_\mathrm{g}} \dfrac{\partial}{\partial r }\left( r v_r \rho_\mathrm{g} C_i \right)  \\  
+  \frac{1}{\rho_\mathrm{g}} \dfrac{\partial}{\partial z }\left( \rho_g D_i \dfrac{\partial C_i}{\partial z}  \right) -  \frac{1}{\rho_\mathrm{g}} \dfrac{\partial}{\partial z }\left(v_z \rho_\mathrm{g} C_i \right),  
\end{split}
\end{equation}
where we have assumed that the diffusion coefficient $D_i$ is the same in the vertical and radial radial direction. Here, subscript $i$ can correspond to CO vapor ($\Cv=\rho_\co/\rho_\mathrm{g}$), small dust ($C_\mathrm{d}=\rho_\mathrm{d}/\rho_\mathrm{g}$), or CO ice present on small dust ($C_\ice=\rho_\mathrm{ice,d}/\rho_\mathrm{g}$), and we will solve Eq.~\ref{eq:dCdt_motion} for all three species. For vapor species, $v_r=v_z=0$ (appropriate for a static disk) and the \new{diffusion coefficient is related to the viscosity through the Schmidt number $\Sc = \nu_\mathrm{T} / D_\mathrm{g}$ for which we will use $\Sc=1$. We return to the assumptions of using a static disk with a constant $\alpha$ in Sect.~\ref{sec:discussion}.}

For dust grains (and the ice present on the grains), radial drift and vertical settling have to be included \citep[e.g.,][]{armitage2010}
\begin{equation}\label{eq:driftsettle}
\begin{split}
v_r =& - 2  \eta r \Omega \frac{ \St }{1+\St^2},\\
v_z =& - \Omega z  \St,
\end{split}
\end{equation}
in which $\eta =  0.5 (c_s/ r \Omega )^2  (\partial \ln \rho_\mathrm{g} / \partial \ln r) \approx (c_s/ r \Omega )^2 \sim 10^{-3}$ represents the dimensionless pressure gradient in the gas disk. The magnitude of the drift and settling velocities depends on the dimensionless Stokes number, a function of particle size $s$ and (material) density $\rho$ through
\begin{equation}
\St = \sqrt{\frac{\pi}{8}} \frac{ s \rho  }{ \rho_\mathrm{g} c_s } \Omega.
\end{equation}
Small, porous dust is then usually well-coupled to the gas\footnote{Except for regions of the disk where the gas density drops significantly, i.e., for $ z>H$ and/or $r> r_c$.}, i.e., $\St \ll 1$ and $v_r \approx v_z \approx 0$, while larger and compact pebbles decouple from the gas and drift and settle significantly \citep{weidenschilling1977}. For solids with a significant Stokes number, the diffusion coefficient deviates from $D_\mathrm{g}$ and is given by $D_\mathrm{d} = D_\mathrm{g}/(1+\St^2)$ \citep{youdin2007}.

\subsection{Freeze-out and desorption of CO}\label{sec:FOandDES}
We combine freeze-out (FO), thermal desorption (TD), and photo-desorption (PD) of CO molecules from/onto grains in a single equation by writing
\begin{equation}\label{eq:dCdt_chem}
\frac{\partial \Cv}{\partial t} =   \frac{3 v_\mathrm{th}}{4 s_\bullet }  \frac{\rho_\mathrm{d}}{\rho_\bullet}   \bigg[     \overbrace{ \frac{\rho_\sat}{\rho_\mathrm{g}} }^{\mathrm{TD}} - \underbrace{ \frac{\rho_\mathrm{v}}{\rho_\mathrm{g}} }_{\mathrm{FO}} +  \overbrace{ \frac{4 m_\co Y F_\uv(z)}{v_\mathrm{th} \rho_\mathrm{g}}  }^{\mathrm{PD}}    \bigg] , 
\end{equation}
where we have assumed that all dust particles contributing to $\rho_\mathrm{d}$ have the same area-to-mass ratio of $3/(s_\bullet \rho_\mathrm{\bullet})$. Conservation of the total amount of CO gives
\begin{equation}
\frac{\partial \Cv}{\partial t} = - \frac{\partial \Cice}{\partial t}.
\end{equation}
The equilibrium vapor density in Eq.~\ref{eq:dCdt_chem} depends on temperature and can be written as
\begin{equation}
\rho_\sat = m_\co  (4 /v_\mathrm{th}) N_s  \times \nu_0 \exp\left\{ - \frac{\mathcal{E}}{ k_\mathrm{B}T} \right\},
\end{equation}
with $\nu_0=( 2 N_s E/\pi^2 m_\co)^{1/2}$, and we use a binding energy\footnote{This value for the binding energy is appropriate for CO-CO binding \citep[see][]{oberg2005}.} $\mathcal{E}/k_\mathrm{B}=850\mathrm{~K}$ and a density of adsorption sites of $N_s=10^{15} \mathrm{~cm^{-2}}$. The thermal velocity is given by $v_\mathrm{th} = \sqrt{8  k_\mathrm{B} T / \pi m_\co }$ with $m_\co = 28 m_\hy$ the mass of a single CO molecule.

The local UV flux is calculated as $F_\uv(z) = F_0 e^{-\tau_\uv(z)}$, with $\tau_\uv$ the integrated vertical depth at height $z$ and the incident flux $F_0 = \Gamma \times G_0$ is defined in terms of the interstellar radiation field $G_0=10^{8}\mathrm{~cm^{-2}~s^{-1}}$. We set $\Gamma=1$. Assuming the UV flux is negligible in the midplane (i.e., $\tau_\mathrm{UV}(z=0) \gg 1$), pebbles lose/gain ice at a rate
\begin{equation}\label{eq:dmicedt}
\frac{\partial m_\ice}{\partial t} = 4\pi s_\mathrm{p}^2 \frac{v_\mathrm{th}}{4} \left( \rho_\mathrm{v} - \rho_\sat \right),
\end{equation}
with $s_\mathrm{p}$ the pebble size.

\subsection{Particle-particle collision velocities}\label{sec:v_rel}
Particle-particle velocities play an important role in determining the outcome and frequency of collisions \citep[e.g.,][]{brauer2007,guttler2010}. \new{We consider 5 sources of relative velocities: Brownian motion $(\Delta v_\mathrm{BM})$, turbulence $(\Delta v_\mathrm{tur})$, and differential settling $(\Delta v_z)$, azimuthal drift $(\Delta v_\phi)$ and radial drift $(\Delta v_r)$, each of which is calculated following \citet[][Sect.~2.3.2]{okuzumi2012}. The different components are then added quadratically}
\begin{equation}\label{eq:v_rel}
v_\mathrm{rel} = \sqrt{ (\Delta v_\mathrm{BM})^2 + (\Delta v_\mathrm{tur})^2 + (\Delta v_z)^2 + (\Delta v_\phi)^2 +  (\Delta v_r)^2}.
\end{equation}

\new{For typical values of $\alpha$ and Stokes numbers $\St<1$, however, the turbulent term is expected to dominate and we have \citep{ormel2007b}}
\begin{equation}\label{eq:v_tur}
v_\mathrm{rel} \approx \Delta v_\mathrm{tur} \approx  \sqrt{\alpha} c_s \times
\begin{cases}
 \RET^{1/4}  \Delta \St &\textrm{~for~}\St<\RET^{-1/2},\\
1.6 \sqrt{\St} &\textrm{~for~}\St>\RET^{-1/2},
\end{cases}
\end{equation}
\new{where $\St$ is the Stokes number of the larger of the two particles, $\Delta \St$ is the difference in Stokes numbers, and the turbulent Reynolds number is the ratio between the turbulent and molecular viscosity $\RET = \nu_\mathrm{T} / \nu_m = (\pi/2)^{1/2} \nu_\mathrm{T} \sigma_\mathrm{mol} \rho_\mathrm{g} / \mu m_\hy c_s$, with $\sigma_\mathrm{mol}=2\times10^{-15}\mathrm{~cm^{-2}}$ the molecular cross section \citep[e.g.,][]{okuzumi2012}.}


\begin{figure}[t]
\centering
\includegraphics[clip=,width=.9\linewidth]{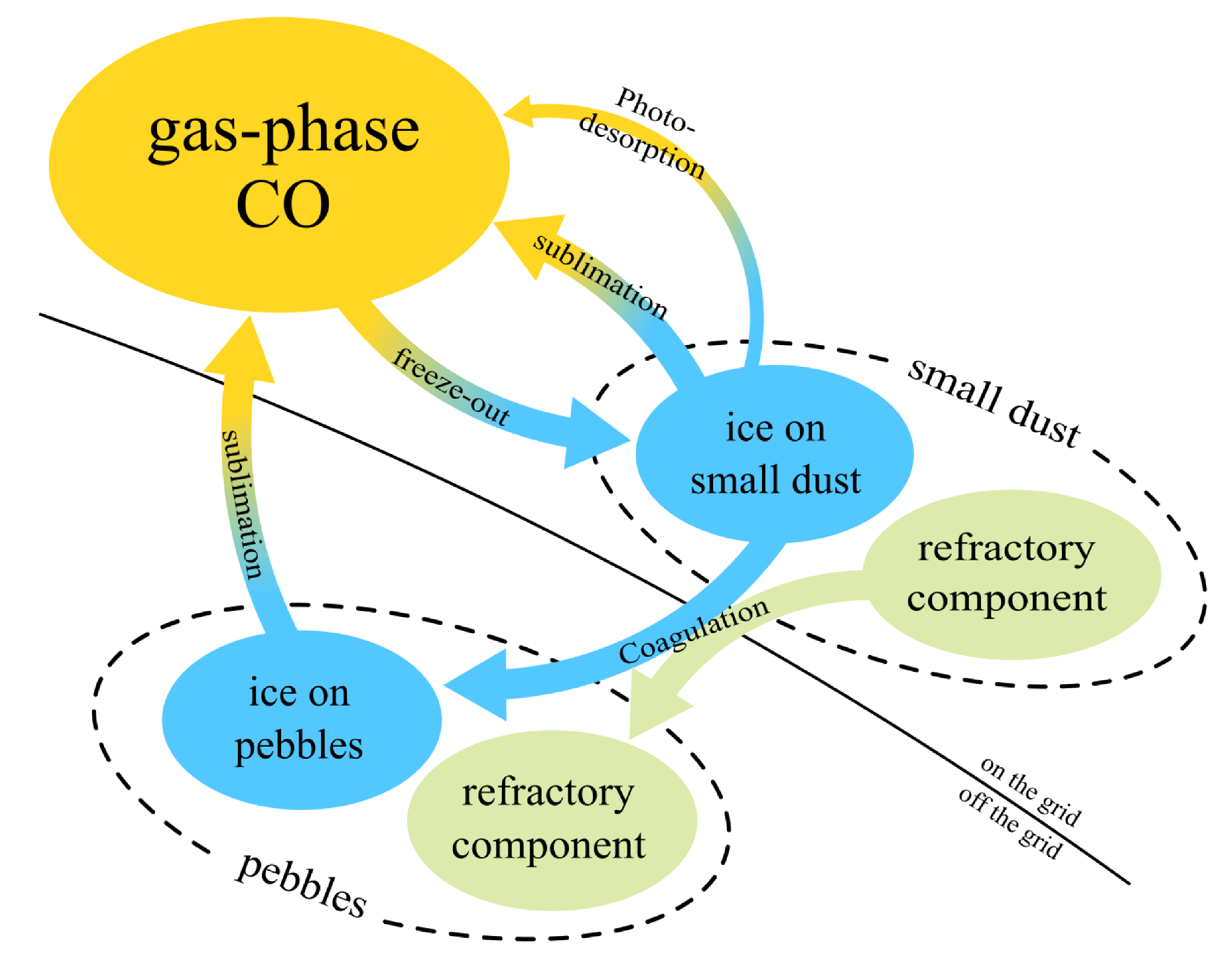}
\caption{Conceptual framework of the hybrid model described in Sect.~\ref{sec:numerical}. The abundances of CO vapor ($\Cv$), small dust grains ($\Cd$), and the ice-frozen-out-on-small-dust-grains ($C_\ice$) are all followed on a logarithmic 2-dimensional grid (Sect.~\ref{sec:grid}). Pebbles, on the other hand, are described using Lagrangian tracer particles, whose motions through the disk are simulated using a Monte Carlo approach (Sect.~\ref{sec:pebbles}). Arrows represent various interactions between different components, all of which are described in more detail in the text.}
\label{fig:sketch}
\end{figure}

\begin{deluxetable}{l  l l }[b]
\centering
\tablecaption{Parameters used throughout this paper.}
\tablewidth{0pt}
\tablehead{
\colhead{Symbol} & \colhead{Description} & \colhead{Values}
}
\startdata
$r_c$ & Disk characteristic radius & $100\mathrm{~au}$\\
$M_\mathrm{disk}$ & Total disk gas mass & $0.05~M_\odot$\\
$p$ & Surface density powerlaw index & $1$\\
$q$ & Temperature powerlaw index & $1/2$\\
$z_q$ & Temperature profile parameter & $3$\\
$\delta$ & Temperature profile parameter & $2$\\
$\Cv^0$ & Initial CO abundance & $0.001$\\
$\Cd^0$ & Initial dust-to-gas ratio & $0.01$\\
$\alpha$ & Turbulence parameter & $10^{-3}$\\
$s_\bullet$ & Monomer size & $0.1\mathrm{~\mu m}$\\
$\rho_\bullet$& Monomer density & $2\mathrm{~g/cm^2}$\\
$\kappa_\mathrm{UV}$ & Monomer opacity & $8\times10^4\mathrm{~cm^2/g}$\\
$Y$ & Photodesorption yield & $10^{-2}/\mathrm{photon}$\\
$N_s$ & adsorption site density & $10^{15}\mathrm{~cm^{-2}}$\\
$f_c$ & timestep parameter & $0.5$\\
$\phi_\mathrm{c}$ & Pebble maximum filling factor& 0.4 \\
$f_\mathrm{eff}$ &Pebble conversion factor & $0.1$\\
$\fw$ & ice stickiness parameter & 0.5 \\
$\mathcal{E}/k_\mathrm{b}$ & CO binding energy & $850\mathrm{~K}$\\
$f_r$ & radial grid spacing & $1.05$\\
$f_z$ & vertical grid spacing & $1.1$\\
$r_0$ & grid inner boundary & $10\mathrm{~au}$
\enddata
\label{tab:params}    
\end{deluxetable}

\begin{figure*}[t]
\centering
\includegraphics[clip=,width=.8\linewidth]{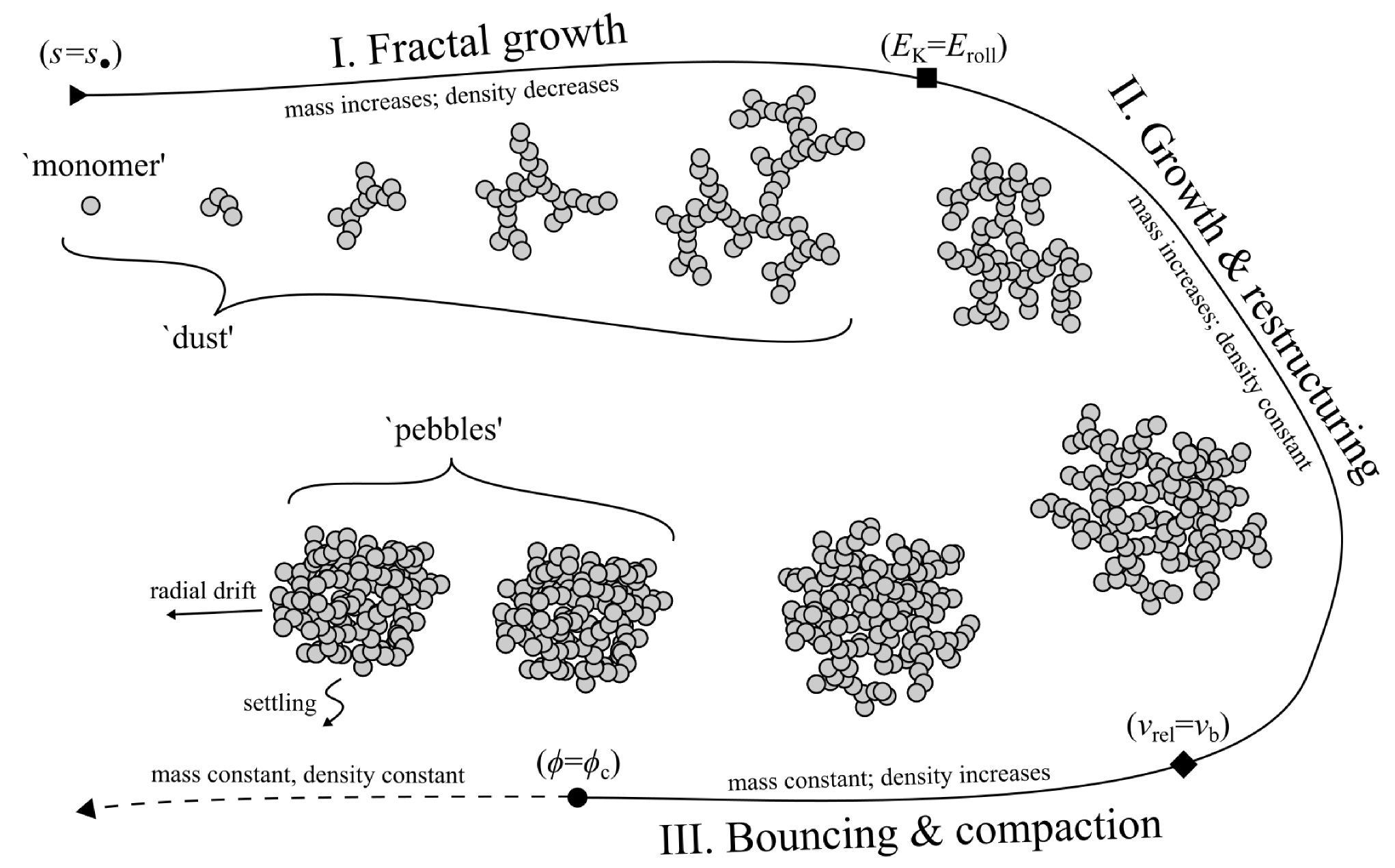}
\caption{Cartoon of how pebble formation proceeds in our model (see Section \ref{sec:localpebbleformation}). Initially, monomers of size $s_\bullet$ grow into dust aggregates with a fractal dimension of ${\approx}2$ (stage I). When the kinetic energy in collisions becomes large enough, restructuring occurs and aggregates grow at a constant internal density (stage II). After the bouncing threshold velocity is exceeded, bouncing collisions efficiently compress the aggregates (stage III). The term `dust' refers to all solids in stage I, and we refer to the end-products of stage III as `pebbles': the end-products of local dust coagulation.}
\label{fig:dust_sketch}
\end{figure*}

\section{Numerical approach}\label{sec:numerical}
The goal of this section is to develop a numerical approach to study, in 2D, the interaction and co-evolution of three distinct components (see Fig.~\ref{fig:sketch}):
\begin{itemize}
\item{\emph{Small dust aggregates:} Composed of sub-micron dust grains, these fractal aggregates are usually well-coupled to the gas and dominate the solid surface area in the protoplanetary disk.}
\item{\emph{Pebbles:} More massive, compacted solids, for which gravitational settling and radial drift are important. The pebble population typically dominates the solid mass in the (inner) disk midplane.}
\item{\emph{CO molecules:} CO molecules can float freely in the gaseous nebula, freeze-out on dust grains (forming CO ice), and end up on pebbles when coagulation takes place. No molecules are created or destroyed in our simulations.}
\end{itemize}
The concentrations of CO vapor, small dust, and ice-on-small-dust are all followed on a 2-dimensional $r{+}z$ grid (Sect.~\ref{sec:grid} and \ref{sec:2dtransport}). The growing population of pebbles, on the other hand, is represented by Lagrangian tracer particles (or, representative particles) and we use a random-walk-like approach (with added settling and radial drift) to track their movement (Sect.~\ref{sec:pebdynamics}). Given some initial conditions (detailed in Sect.~\ref{sec:IC}), we calculate forward in time using a combination of implicit and explicit techniques to account for the interactions shown in Fig.~\ref{fig:sketch}.

\subsection{Grid \& boundary conditions}\label{sec:grid}
Similar to \citet{ciesla2009}, we set up a logarithmic grid with $r_{i+1}/r_i=1.05$ and $z_{i+1}/z_i=1.1$, with $r_0=10\mathrm{~au}$ and $z_0=0.25\mathrm{~au}$ (at every radius). The number of cells in the radial direction is chosen such that the outer radius corresponds to approximately $3r_c = 300\mathrm{~au}$. Every grid cell can be thought of as a ring with volume $\mathcal{V} = 2\pi r \Delta r \Delta z$, covering an area $\mathcal{A}=2\pi r \Delta r$ when projected on to the midplane\footnote{Note that $\mathcal{V}$ and $\mathcal{A}$ vary significantly between cells.}. \new{The boundary conditions are reflective at the midplane ($\partial C / \partial z =0$) and at the inner and outer boundary of the domain ($\partial C / \partial r =0$). While this means no dust is lost through diffusion through the inner boundary, radially drifting pebbles can be lost to the inner disk (Sect.~\ref{sec:pebdynamics}). Transport in the disk is limited to $z/H\leq4$ by setting the diffusivities and initial concentrations to $0$ above $z/H=4$ (similar to \citealt{ciesla2009})}.

\subsection{Initial conditions}\label{sec:IC}
At $t=0$, we start out with well-mixed dust and CO vapor: $\Cd(r,z)=\Cd^0=10^{-2}$ and $\Cv(r,z)=\Cv^0 = 10^{-3}$ (corresponding to roughly $10^{-4}$ CO molecules per $\hyy$ molecule). Then, we allow the CO to freeze-out until an equilibrium is reached in every grid cell. No pebbles exist at the start of the calculations.

\subsection{Transport of vapor, dust, and ice}\label{sec:2dtransport}
Following \citet{ciesla2009}, the transport of vapor, ice, and small dust is calculated by explicit integration of Eq.~\ref{eq:dCdt_motion} for each component using the method of finite differences with a time step chosen as a fraction $f_c$ of the smallest (vertical) diffusion timescale across any grid cell:
\begin{equation}\label{eq:delta_t}
\Delta t = f_c \times \min \left\{ \frac{ (\Delta z_i)^2}{D_{\mathrm{g},i}}  \right\},
\end{equation} 
where we use $f_c = 0.5$. For our disk model and grid set-up, the rhs of Eq.~\ref{eq:delta_t} is usually dominated by the midplane cell at the outer edge of the disk, because $\Delta z / H$ decreases with radius for grid cells of a fixed vertical size.

\subsection{Interaction between vapor and dust}
\new{The interaction between CO molecules and dust grains} is solved implicitly: in the rhs of Eq.~\ref{eq:dCdt_chem}, only the 2nd term depends on $\Cv$. Thus, by defining $C^*\equiv \Cv - \rho_\sat/\rho_\mathrm{g} - 4m_\co Y F_\uv(z) / v_\mathrm{th} \rho_\mathrm{g} $, we can rewrite
\begin{equation}\label{eq:freezeout}
\frac{\partial C^*}{\partial t} = - \underbrace{\frac{3}{4} \frac{v_\mathrm{th} \rho_\mathrm{d} }{s_\bullet \rho_\mathrm{\bullet}}}_{\equiv A_\mathrm{ch}} C^* ,
\end{equation}
so that $C^*(t+\Delta t) /C^*(t)= 1 - \exp( -A_\mathrm{ch} \Delta t )$, where an additional constraint comes from $\Delta C^* \leq C_\ice$, i.e., there is only so much ice that can be released. When condensing vapor is added to the small dust grains in the form of ice, we assume the formation of the ice mantle has a negligible effect on the size, mass, and Stokes number of the small dust. At the start of the simulation this is reasonable because $\Cv/\Cd=0.1$, so CO ice can contribute at most 10\% to a particle's mass. However, in specific regions of the midplane the ice fraction of small grains can become substantially larger towards the end of the simulation, an effect we describe in Sect.~\ref{sec:driftingpebbles}.

\subsection{Pebble formation}\label{sec:pebbles}
The purpose of this Section is to develop a frame-work that allows us to convert microscopic dust into pebbles in our simulations in a simplified, but physically motivated way. To that end, we first discuss how local dust coagulation is believed to proceed and what the end-products (the `pebbles') are. Then, we describe how the conversion of dust to pebbles is handled in our numerical model.

\subsubsection{Local dust coagulation}\label{sec:localpebbleformation}
The smallest grains in our simulation are monomers with radius $s_\bullet$ and material density $\rho_\bullet$. On timescales of hundreds to thousands of orbital periods, these grains will coagulate into larger aggregates \citep[e.g.,][]{dominik2006}. Dust coagulation can be split up into three stages, depicted in Fig.~\ref{fig:dust_sketch}:

\emph{Stage I: Fractal growth.} As monomer grains coagulate at initially low velocities, it is expected that very porous, fractal structures form with a fractal dimension close to 2 \citep{wurm1998,kempf1999}, which means their surface-area-to-mass ratio stays constant. This fractal growth phase lasts until the kinetic energy in collisions exceeds a threshold energy $E_\mathrm{roll}$, the energy needed to restructure monomer-monomer bonds \citep{dominiktielens1997}. The rolling energy depends on the material properties of the monomer surface, and is expected to be larger for surfaces dominated by water ice (Sect.~\ref{sec:stickiness}). \new{Following \citet{okuzumi2012}, we compare the kinetic energy in aggregate-aggregate collisions to the rolling energy and obtain the critical aggregate mass at which restructuring starts to occur
\begin{equation}\label{eq:m_roll}
\begin{split}
m_\mathrm{roll} &= \dfrac{32}{\pi} \dfrac{E_\mathrm{roll}}{\alpha \sqrt{\RET}} \rho_\mathrm{g}^2 \left( \Omega s_\bullet \rho_\bullet \right)^{-2},\\
&= \dfrac{32}{\pi} \left(\dfrac{2}{\pi} \right)^{1/4} \frac{E_\mathrm{roll}}{\alpha^{3/2}} \left( s_\bullet \rho_\bullet \right)^{-2} \sqrt{\dfrac{m_\mathrm{g} \Omega}{\sigma_\mathrm{mol} \rho_\mathrm{g} c_s}} \left( \dfrac{\rho_\mathrm{g}}{\Omega} \right)^{2},
\end{split}
\end{equation}
where we have used that fractal aggregates are in the Epstein drag regime and approximated $\Delta \St \approx \St$ in the first regime of Eq.~\ref{eq:v_tur}.}

\begin{figure*}[t!]
\centering
\includegraphics[clip=,width=1.\linewidth]{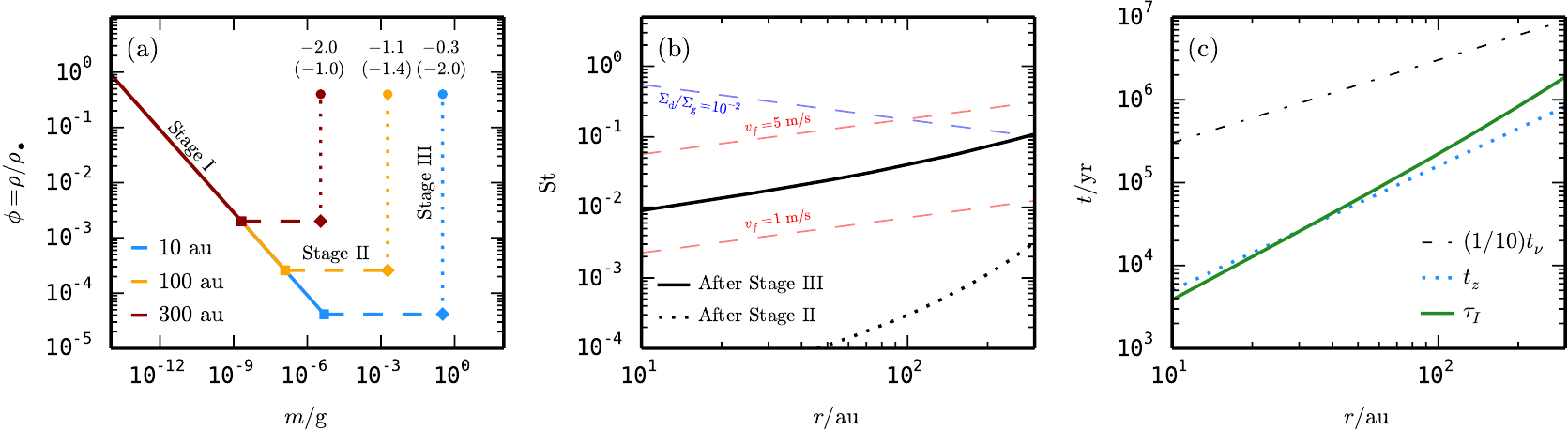}
\caption{\new{Illustration of the pebble formation framework (Sect.~\ref{sec:pebbles}) for the disk model of Sect.~\ref{sec:diskmodel} and using $s_\bullet=0.1\mathrm{~\mu m}$, $\fw=0.5$, and $\phi_\mathrm{c}=0.4$. (a): The three stages of local dust growth (see Sect. \ref{sec:localpebbleformation}) and their end-products at three different midplane locations. Numbers correspond to $\log_{10}(s/\mathrm{cm})$ and $\log_{10}(\St)$ (in brackets). (b): Stokes numbers of pebbles (solid black curve) formed in the midplane as a function of heliocentric distance. The red dashed curves depict a constant fragmentation threshold $v_f$ \citep[][Eq.~3]{birnstiel2012} of $1$ and $5\mathrm{~m/s}$, and the blue dashed curve shows the maximum Stokes number in the drift-limited scenario \citep[][Eq.~17]{birnstiel2012} for a dust-to-gas ratio of $10^{-2}$. (c): Timescale for pebble formation in the midplane from Eq.~\ref{eq:tau_I} (green curve) compared to the vertical mixing timescale $\tau_z\sim(\alpha \Omega)^{-1}$ and the viscous timescale $t_\nu$ for $\alpha=10^{-3}$.}}
\label{fig:pbf}
\end{figure*}

\emph{Stage II: Growth at a constant porosity.} Aggregates more massive than $m_\mathrm{roll}$ will be compacted in collisions and their porosity is not expected to increase any further. Instead, as long as sticking is common in aggregate-aggregate collisions, further growth takes place at a nearly constant internal density \citep{okuzumi2012,kataoka2013c}. During this growth phase, however, the Stokes number (and therefore collision velocities) increase and perfect sticking in collisions is no longer guaranteed \citep[e.g.,][]{blumwurm2000}. According to \citet{guttler2010,weidling2012}, the probability of a collision resulting in sticking will decrease with increasing particle mass and collision velocity and thus we can define, for a given particle mass (and composition), a critical threshold velocity $v_\mathrm{b}$ above which the probability of sticking is zero (Sect.~\ref{sec:stickiness}). To determine how long Stage II can proceed, we incrementally increase $m$ starting from $m_\mathrm{roll}$, updating the relative collision velocity\footnote{\new{When calculating the relative velocity, we combine all 5 velocity sources listed in Sect.~\ref{sec:v_rel} and use $\Delta \St \approx \St$ for terms that rely on the difference in Stokes numbers.}} and critical bouncing threshold $v_\mathrm{b}$ as the aggregate's mass increases (while keeping the internal density constant). The point at which $v_\mathrm{rel} = v_\mathrm{b}$ then marks the end of Stage II.

\emph{Stage III: Compaction in bouncing collisions.} During Stage II, aggregates maintain a fairly high porosity, with internal densities $\rho/\rho_\bullet \sim 10^{-5}-10^{-3}$ \citep[e.g.,][]{okuzumi2012,kataoka2013c,krijt2015}. During the final stage of local dust evolution, we imagine the frequent bouncing collisions act to compress the aggregates and increase their internal density (while keeping their mass constant), effectively decreasing their size and increasing their Stokes number. Compaction in successive bouncing collisions is a complex process and has only been studied experimentally for a narrow range of (initial) porosities, collision velocities, and materials \citep[e.g.,][]{weidling2009,guttler2010}. Here, we simply assume compression increases the aggregate density to $\rho=\phi_\mathrm{c}\rho_\bullet$ and we treat the $\phi_\mathrm{c}\leq1$ as a free parameter. The compacted aggregates formed at the end of Stage III are called pebbles.

Figure~\ref{fig:pbf}(a) shows the three stages described above at three different locations in the disk. At smaller radii, more extreme porosities are reached at the end of stage I \citep[see also][Fig.~10]{okuzumi2012}, and while pebbles in the inner disk are born with bigger physical sizes, their Stokes numbers are actually smaller than those of pebbles forming in the outer disk. In Fig.~\ref{fig:pbf}(b), we show the Stokes numbers of pebbles that are created at different radii in the disk, and compare them to the Stokes numbers at the end of Stage II, and to maximum Stokes numbers expected from fragmentation-limited as well as drift-limited growth \citep{birnstiel2012}. We see that the Stokes number increases considerably during Stage III (the compaction stage), \new{with the final Stokes numbers falling below the fragmentation limit for $v_f=5\mathrm{~m/s}$ and the drift limit for $\Sigma_\mathrm{d}/\Sigma_\mathrm{g}=0.01$.}

\subsubsection{Rolling energy and bouncing treshold}\label{sec:stickiness}
Based on experimental results of \citet{heim1999,gundlach2011} in combination with the contact theory of \citet{krijt2014}, the rolling energy can be obtained as
\begin{eqnarray}
E_\mathrm{roll}^\water&=1.4\times10^{-7} \mathrm{erg} \,(s_\bullet/\mathrm{\mu m})^{5/3} &\textrm{~~for $\water$ ice,} \nonumber \vspace{3mm} \\
E_\mathrm{roll}^\sil&=2.3\times10^{-8} \mathrm{erg} \,(s_\bullet/\mathrm{\mu m})^{5/3} &\textrm{~~for dust,}
\end{eqnarray}
for water ice and dust. For a collection of monomers whose surface properties are a mix between those of water-ice and non-water-ice, the characteristic rolling energy can be interpolated as \citep{lorek2016,lorek2017}
\begin{equation}
E_\mathrm{roll} = \fw E_\mathrm{roll}^\water + (1-\fw)E_\mathrm{roll}^\sil,
\end{equation}
where $0 \leq \fw \leq 1$ is a fraction indicating how dominant water ice is.

For collisions between similar dust aggregates, the threshold velocity above which sticking is no longer possible ($v_b$) has been experimentally constrained by \citet{guttler2010,weidling2012} as
\begin{equation}\label{eq:v_thr}
v_\mathrm{b}^\sil = (m / 3.3\times10^{-3}\mathrm{~g})^{-5/18} \mathrm{~cm/s}.
\end{equation}
The threshold velocities for water-ice aggregates are expected to be a factor $10$ larger \citep{wada2013,gundlach2014}, while $\coo$ ice grains behave more like bare silicate dust grains \citep{musiolik2016,musiolik2016b}. Here we use the behavior of $\coo$-ice as a proxy for those covered with $\co$. We again follow \citet{lorek2016,lorek2017} and interpolate the threshold velocities for a mixed material as
\begin{equation}
v_\mathrm{b} = \fw v_\mathrm{b}^\water + (1-\fw) v_\mathrm{b}^\sil = (1+9\fw ) v_\mathrm{b}^\sil.
\end{equation}

In terms of the dust behavior, the main free parameters are then: the monomer size ($s_\bullet$), the degree of collisional compaction in bouncing collisions (captured in $\phi_\mathrm{c}$), and the extent to which the surface properties of the monomers are dominated by water ice ($\fw$). This last parameter plays a role in determining the rolling energy (i.e., how readily is an aggregate compressed) and in setting the transition from sticking to bouncing. For core/mantle grains, the fraction $\fw$ cannot be directly equated to the water ice mass faction of a grain because the structure/layering of the ice mantle matters as well: even a monomer whose mass is dominated by $\water$-ice can have $\fw\sim0$ if its surface consists of $\co$ or $\coo$ ice. We opt for using a constant $\fw$ during our simulations, and study the sensitivity of the results on the choice for $\fw$ in Sect.~\ref{sec:results}.

\subsubsection{Converting dust into pebbles}\label{sec:pebbleconversion}
Instead of following the incremental growth from monomers to pebbles \citep[e.g.,][]{ormel2007,zsom2010,krijt2016}, we opt for a more stochastic approach in which grid-cells occasionally convert a fraction $f_\mathrm{eff}$ of their dust content into pebbles, creating Lagrangian tracer particles when this occurs. To determine whether pebbles are created during a given timestep, we first estimate how long the growth process from monomer to pebble is expected to take at that location, \new{and then use that timescale in combination with a random number to decide wether pebbles are formed or not.}

\new{In vertically integrated models that focus on compact (i.e., non-fractal) grains, the mass doubling timescale associated with pebble formation is often written as $\tau_m \approx (\Sigma_\mathrm{g}/\Sigma_\mathrm{d}) / \Omega$ \citep[e.g.,][]{birnstiel2012,lambrechts2014,drazkowska2016,booth2017}, assuming growing particles have Stokes numbers that are large enough for them to settle to the midplane and to have relative velocities that scale with $v_\mathrm{rel}\propto \sqrt{\St}$ (see Eq.~\ref{eq:v_tur}), in which case the growth timescale in the midplane becomes independent of particle size and the value of $\alpha$. In our picture, however, the initial fractal growth phase (Stage I) results in the aggregate's Stokes number staying small for a large range of masses (Fig.~\ref{fig:pbf}(a)). During this initial phase, for particles with masses $m < m_\mathrm{roll}$, the mass-doubling timescale due to collisions with like-size aggregates can be obtained as}
\begin{equation}\label{eq:tau_m}
\tau_m \equiv \frac{ m }{(\partial m / \partial t)} \approx \frac{m}{\sigma_\mathrm{col} \rho_\mathrm{d} v_\mathrm{rel}} \approx \sqrt{\frac{8}{\pi}} \frac{\rho_\mathrm{g}}{\rho_\mathrm{d} \sqrt{\alpha} \Omega \RET^{1/4}},
\end{equation}
\new{again making use of Eq.~\ref{eq:v_tur}. Since this timescale does not depend on $m$, the total time it takes for monomers to grow into aggregates with mass $m_\mathrm{roll}$ can be estimated as $\tau_m$ multiplied by the number of times a monomer's mass needs to double, i.e.,}
\begin{equation}\label{eq:tau_I}
\tau_I = \tau_m \log_{2}\left( \frac{m_\mathrm{roll}}{m_\bullet} \right).
\end{equation}
\new{Here, we approximate\footnote{\new{This approximation ignores that for the smallest of grains Brownian motion leads to growth timescales that are shorter than Eq.~\ref{eq:tau_m} \citep[e.g.,][]{zsom2010,krijt2015} because turbulence will dominate relative velocities for the majority of Stage I. In addition, while we assume that the duration of Stage III is short, this phase could well last a significant amount of time; in particular when $\phi_\mathrm{c} \gtrsim 0.2$, as many collisions can be needed to reach such high filling factors \citep{weidling2012}.}} the total time it takes to grow from a monomer to a compact pebble as being dominated by Stage I and given by Eq.~\ref{eq:tau_I}. Figure~\ref{fig:pbf}(c) shows $\tau_I$ in the midplane as a function of heliocentric distance assuming the initial $\rho_\mathrm{d}/\rho_\mathrm{g}=0.01$; the conversion of dust into pebbles takes ${\sim}10^{3-4}\mathrm{~yr}$ at 10 au and ${>}10^{6}\mathrm{~yr}$ outside of $r=100\mathrm{~au}$.} 

Now that we have an idea of how long coagulation is expected to take, we generate a random number $R_0$ between $(0,1]$ which we use together with $\tau_I$ and the duration of the timestep $\Delta t$ to \new{determine whether a fraction $f_\mathrm{eff} \leq 1$ of dust is converted to pebbles:}
\begin{equation}\label{eq:F_ice}
\begin{cases}
-\ln(R_0) > \Delta t / ( f_\mathrm{eff} \tau_I): &\textrm{no pebbles created,}  \vspace{3mm} \\
-\ln(R_0) \leq \Delta t / (f_\mathrm{eff} \tau_I): &\textrm{$f_\mathrm{eff} \rho_\mathrm{d}$ converted to pebbles.}
\end{cases}
\end{equation}
The pebble formation timescale is a function of the dust abundance inside the cell, $\tau_I \propto (\rho_\mathrm{d}/\rho_\mathrm{g})^{-1}$, so pebble creation becomes increasingly unlikely as dust is removed. \new{In addition, we do not allow pebbles to form in cells for which the relative velocity between monomers exceeds $1\mathrm{~m/s}$.} When pebbles are formed, we create a tracer particle that we place inside the appropriate grid cell. The size of the newly-formed pebbles is given by the paradigm outlined in Section \ref{sec:localpebbleformation}. The total mass the tracer represents is $\mathcal{M}= f_\mathrm{eff} \rho_\mathrm{d} \mathcal{V}$, so the number of represented particles equals $\mathcal{N}=\mathcal{M}/m_\mathrm{p}$, with $m_\mathrm{p}=(4/3)\pi  \phi_\mathrm{c} \rho_\bullet s_\mathrm{p}^3$ the mass of an individual pebble. Finally, the formed pebbles have an CO-ice-to-rock ratio that reflects that of the small dust (i.e., $\Cice/\Cd$) at the time and location of their formation. \new{We set $f_\mathrm{eff}=0.1$, which typically results in the formation of ${\sim}10^{4-5}$ tracer particles in the simulations presented in this study.}

An advantage of using tracers to represent the pebble population is that this approach allows us to keep track of each individual tracer's history/trajectory as it moves through the disk, while also allowing pebbles of different sizes, make-ups and histories to be present in the same location of the disk. Such information can be used to track the provenance and detailed evolution of particles that er found at a given location in the disk.

\subsection{Pebble dynamics}\label{sec:pebdynamics}
Once pebbles form, we calculate their motions through the disk using the methodology outlined in \citet{ciesla2010,ciesla2011}\footnote{This approach assumes pebbles are always in the strong coupling limit of  \citet{ormel2018}. For pebbles with Stokes numbers $\St \lesssim 0.1$ this approximation is justified.}
\begin{eqnarray}
r(t+\Delta t) =& r(t) + v_r^\mathrm{eff}\Delta t + R_1 \left( \dfrac{2}{\xi} D_\mathrm{p} \Delta t   \right)^{1/2}  ,\label{eq:r_t} \\
z(t+\Delta t) =& z(t) + v_z^\mathrm{eff}\Delta t +R_2 \left( \dfrac{2}{\xi} D_\mathrm{p} \Delta t   \right)^{1/2},
\end{eqnarray}
where $D_\mathrm{p} = D_\mathrm{g} / (1+\mathrm{St_p^2})$ is the pebble diffusivity calculated at $(r(t),z(t))$, $R_1$ and $R_2$ are random numbers between $[-1,1]$ and $\xi=1/3$. The effective velocities are given by
\begin{eqnarray}\label{eq:v_eff}
 v_r^\mathrm{eff} =& v_r & +  \dfrac{\partial D_\mathrm{p}}{\partial r} + \dfrac{ D_\mathrm{p}}{\rho_\mathrm{g}} \dfrac{\partial \rho_\mathrm{g}}{\partial r} ,\\
 v_z^\mathrm{eff}=& v_z &  + \dfrac{ D_\mathrm{p}}{\rho_\mathrm{g}} \dfrac{\partial \rho_\mathrm{g}}{\partial z},
\end{eqnarray}
where we have set $\partial D_\mathrm{p} / \partial z=0$. The drift and settling velocities are given by Eq.~\ref{eq:driftsettle}. We use the same boundary conditions for the pebbles as for the small dust: reflective at $z=0$ and at the top of the grid, and open at the inner and outer disk edges. In practice, most pebbles will eventually leave the grid by drifting through the inner boundary at $r=10\mathrm{~au}$. When this occurs, we remove them from the simulation after recording their properties and the time at which they reached the inner edge.

\subsection{Pebble sublimation}
When an ice-rich tracer particle drifts into an environment where the ice is expected to sublimate, the pebble ice content is evolved using Eq.~\ref{eq:dmicedt} and connected to the vapor density in the cell it resides in through
\begin{equation}
\frac{\partial \Cv}{\partial t} = - \frac{1}{\mathcal{V} \rho_\mathrm{g}} \sum_i \left(\mathcal{N} \frac{\partial m_\ice}{\partial t} \right)_i,
\end{equation}
where the sum is over all super-pebbles that are shedding ice in that same grid cell. Because the small fractal dust will generally dominate the surface area, we ignore direct freeze-out of CO onto pebbles.

\begin{figure*}[t]
\centering
\includegraphics[clip=,width=1\linewidth]{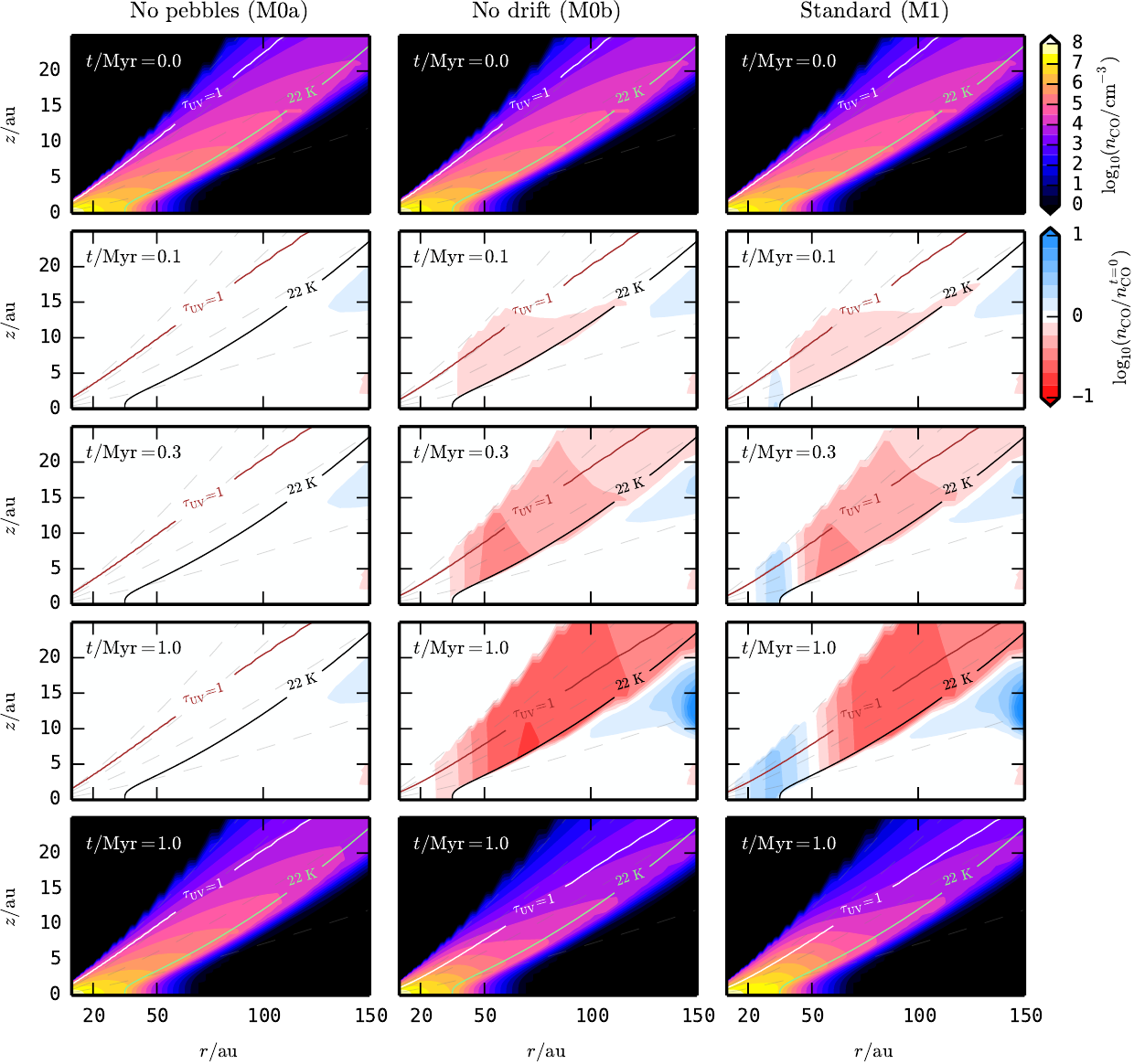}
\caption{Time evolution (top-to-bottom) of the gas-phase CO number density in models M0a, M0b, and M1 (left-to-right). Top and bottom rows show number densities while the middle three rows show number density relative to the initial conditions at $t=0$, with blue representing an enhancement and red a depletion in gas-phase CO density. Contours for $\tau_\uv=1$ and $T=22\mathrm{~K}$ are also drawn and the faint dashed lines indicate $z/H=\{1,2,3,4\}.$}
\label{fig:2D_gas}
\end{figure*}

\begin{figure*}[]
\centering
\includegraphics[clip=,width=1\linewidth]{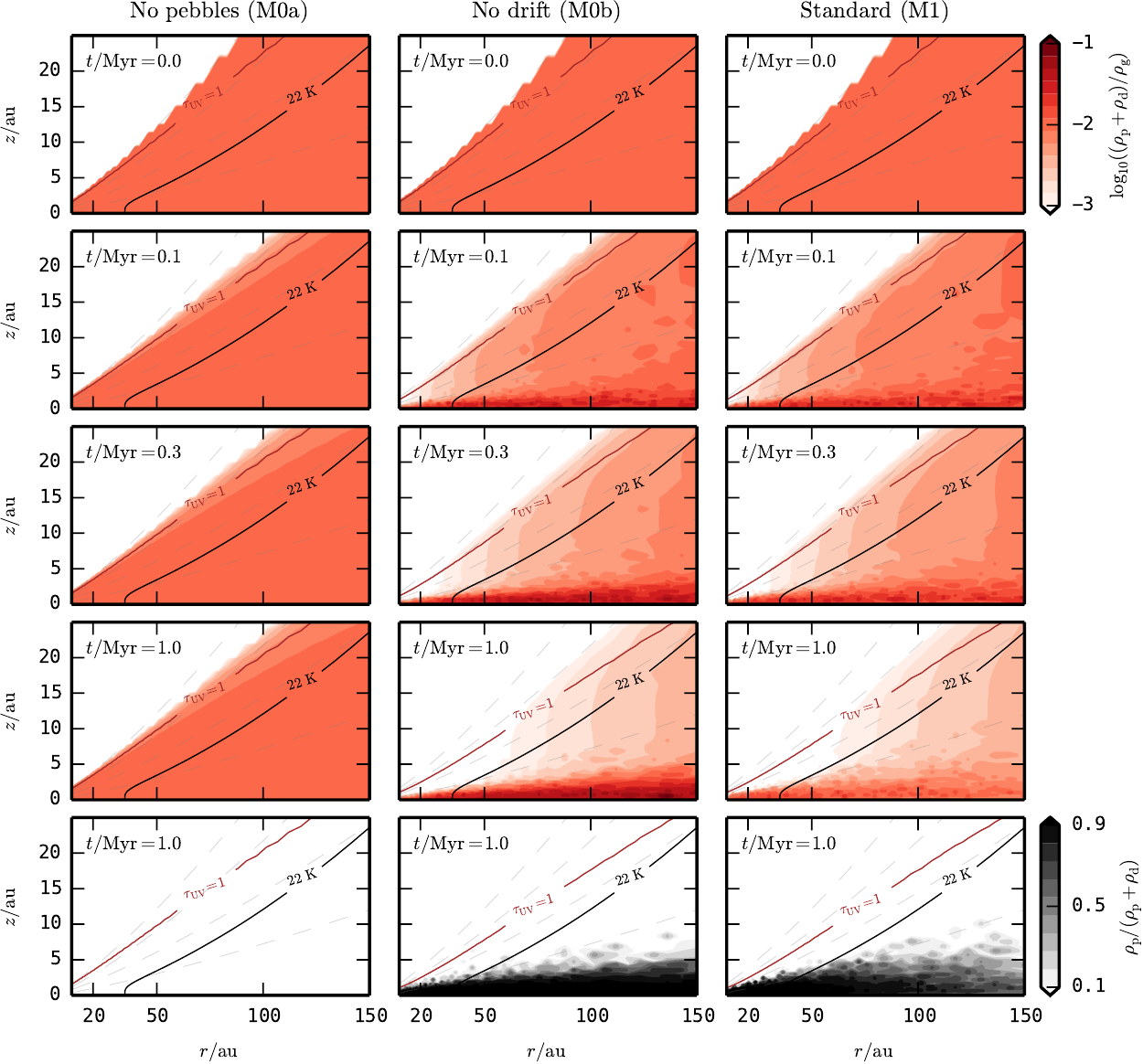}
\caption{Time evolution (top-to-bottom) of the solids in models M0a, M0b, and M1 (left-to-right). The top four rows show the solid-to-gas ratio, i.e., $(\rho_\mathrm{p}+\rho_\mathrm{d})/\rho_\mathrm{gas}$, excluding the contribution of CO ice, while the bottom row shows to what extent pebbles dominate the solid budget locally. Contours for $\tau_\uv=1$ and $T=22\mathrm{~K}$ are also drawn and the faint dashed lines indicate $z/H=\{1,2,3,4\}.$}
\label{fig:2D_sol}
\end{figure*}

\begin{figure*}[]
\centering
\includegraphics[clip=,width=1\linewidth]{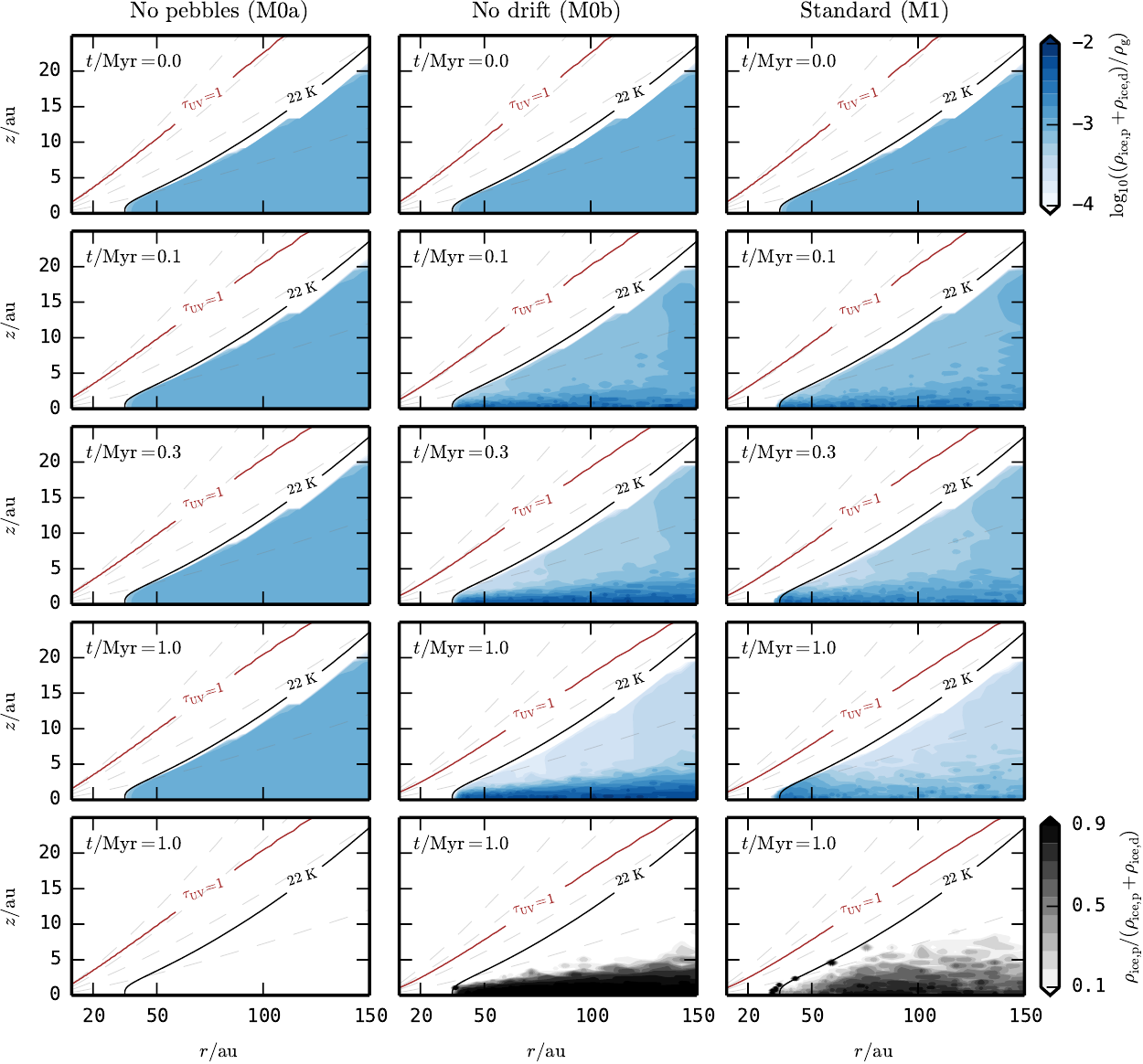}
\caption{Time evolution (top-to-bottom) of the CO ice in models M0a, M0b, and M1 (left-to-right). The top four rows show the ice-to-gas ratio, i.e., $(\rho_\mathrm{ice,p}+\rho_\mathrm{ice,d})/\rho_\mathrm{gas}$, excluding the contribution of CO ice, while the bottom row shows to what extent pebbles dominate the ice budget locally. Contours for $\tau_\uv=1$ and $T=22\mathrm{~K}$ are also drawn and the faint dashed lines indicate $z/H=\{1,2,3,4\}.$}
\label{fig:2D_ice}
\end{figure*}


\section{Results}\label{sec:building}
We first look at simulations of increasing complexity with the goal of understanding how different processes can impact the (re)distribution of CO throughout the nebula and then explore the impact of changing various main parameters in the complete model in Sect.~\ref{sec:results}. The processes that are included in subsequent simulations are summarized in Table~\ref{tab:modelruns} and simulations will typically span a period of $1\mathrm{~Myr}$. Time-series of the distributions of gas-phase CO, the solids and the CO ice are presented in Figs.~\ref{fig:2D_gas}--\ref{fig:2D_ice} and each model is discussed in detail below.

\subsection{No pebble formation}\label{sec:M0a}
In the first scenario, model M0a, dust coagulation (and therefore pebble formation) is not included: dust particles are always and everywhere assumed to be monomers or small fractal aggregates that behave identically to monomers. Over the course of $1\mathrm{~Myr}$, very little change is observed in the gas-phase CO distribution (left column of Fig.~\ref{fig:2D_gas}). The main reason for this is that while CO molecules are continuously being transported vertically and radially, the efficiencies of the processes governing this transport (turbulent diffusion) are essentially identical for gas-phase CO and for CO molecules that are frozen out on small grains, leading to insignifcant net fluxes of CO molecules. 

This situation can be compared to the work of \citet{xu2016}, who used 1D vertical models to study mixing of CO vapor and ice in situations without grain growth. For a grain size of $0.1\mathrm{~\mu m}$ and a vertically constant $\alpha$, \citet{xu2016} find that getting significant CO depletion is only possible at radii at which grains start to decouple from the gas at heights that are comparable to, or below, the location of the surface snowline (see their their Fig.~4 and Sect.~3.2). Thus, significant depletion was hard to achieve for sub-micrometer grains and/or high values of $\alpha$. These findings are supported by our simulations: the decoupling of $0.1\mathrm{~\mu m}$ grains from the gas in our model M0a happens far above the $T=22\mathrm{~K}$ contour (see Fig.~\ref{fig:2D_sol}), and therefore does not result in a depletion of CO vapor from the upper regions. Outside of $r=150\mathrm{~au}$, the stratification in the small-dust distribution gets closer to the surface snowline however, and the warm CO vapor becomes depleted by several 10s of \%, this can be seen in Fig.~\ref{fig:snapshots1}(a), where we have plotted radial profiles of the situation after 1~Myr for the models presented in this Section. For larger monomers and/or lower values of $\alpha$, grains will decouple at smaller $z/H$ \citep[e.g.,][]{dullemond2004}, making CO depletion in the case without coagulation possilbe \citep[see][Fig.~4]{xu2016}.

When CO vapor is mixed down into the $T\lesssim22\mathrm{~K}$ region, it will freeze out on a timescale that depends on the temperature of the gas and the amount of solid surface area that is available \citep[][Sect.~3.2]{bergin2014}. Identifying the freeze-out timescale as $\tau_\mathrm{fo}\approx A_\mathrm{chem}^{-1}$ (see Eq.~\ref{eq:freezeout}), we see that it is proportional to $s_\bullet \rho_\bullet$ (the inverse of the surface-area-to-mass ratio of the dust grains), and inversely proportional to $\rho_\mathrm{d}$ and $v_\mathrm{th}\propto T^{1/2}$, i.e., freeze-out takes longer when grains are bigger, the dust density is lower, and/or the temperature is lower. If the freeze-out timescale is comparable to the (local) radial or vertical transport timescales, vapor molecules that are mixed down can travel significant distances before freezing-out \citep[e.g.,][]{monga2015}, increasing the vapor abundance in these cold regions when compared to the initial (equilibrium) conditions. In model M0a, the freeze-out timescale is generally quite short (fractal aggregates have a large area-to-mass ratio and $\rho_\mathrm{d}\approx 0.01\rho_\mathrm{g}$ everywhere) and CO molecules freeze-out close to the 22~K contour. Around/outside $r=150\mathrm{~au}$ and $z/H>1$, however, the low gas densities increase $\tau_\mathrm{fo}$ somewhat, resulting in slightly elevated gas-phase CO abundances (visible as a blue blob in the middle three rows of Fig.~\ref{fig:2D_gas}), even though the physical number density of CO (i.e., not scaled to the initial value) in these regions is very small (see bottom row of Fig.~\ref{fig:2D_gas}).

\begin{deluxetable*}{r  c c | c | c c c }[t]
\centering
\tablecaption{Parameters used in different model runs.}
\tablewidth{0pt}
\tablehead{
Model ID $\rightarrow$ & \colhead{M0a} & \colhead{M0b}  & \colhead{M1}  & \colhead{M2a} & \colhead{M2b} & \colhead{M2c} 
}
\startdata
Vapor diffusion 					& \checkmark 	& \checkmark 	& \checkmark 	& \checkmark	& \checkmark	& \checkmark \\
Freeze-out/sublimation 			& \checkmark 	& \checkmark 	& \checkmark 	& \checkmark	& \checkmark	& \checkmark \\
Dust and ice dynamics 				& \checkmark 	& \checkmark 	& \checkmark 	& \checkmark	& \checkmark	& \checkmark \\
Pebble formation 				& $\times$ 	& \checkmark 	&\checkmark 	& \checkmark	& \checkmark	& \checkmark \\
Pebble settling					& $\times$ 	& \checkmark 	&\checkmark 	& \checkmark   & \checkmark	& \checkmark \\
Pebble radial drift				& $\times$ 	& $\times$ 	& \checkmark 	& \checkmark   & \checkmark	& \checkmark \\
		 					&  & &  & & &   \\
$\phi_\mathrm{c}$				& 0.4 & 0.4 & 0.4 &  0.04 & - &- \\		
$\fw$							& 0.5 & 0.5 & 0.5& - &  1.0  &  -\\
$\alpha$						& $10^{-3}$ & $10^{-3}$ & $10^{-3}$ & - & - & $10^{-4}$ 
\enddata
\tablecomments{In the lower half of the table, entries that are not shown default to the parameters of models M1.}
\label{tab:modelruns}    
\end{deluxetable*}\vspace{1cm}

\subsection{Non-drifting pebbles}
In model M0b, dust coagulation is included, but the pebbles that form are only allowed to settle vertically, \emph{not} to drift radially. Dust coagulation and the subsequent vertical settling of pebbles results in a dense midplane layer of solids, depleting the dust content in the upper layers (Fig.~\ref{fig:2D_sol}). Since dust evolution is faster at smaller radii \citep[e.g.,][or Fig.~\ref{fig:pbf}(c)]{krijt2016}, the depletion of small dust proceeds from the inside out. After a million years, the small dust density has decreased by about an order of magnitude inside 100 au and the solid mass in the midplane is dominated by pebbles at all radii (see bottom row of Fig.~\ref{fig:2D_sol}).

A consequence of concentrating solids in the disk midplane is that a large fraction of the CO ice will be sequestered there as well. Comparing models M0a and M0b in Fig.~\ref{fig:2D_ice}, it is indeed clear that a large fraction of the CO ice at radii ${>}35\mathrm{~au}$ resides on settled pebbles that are (virtually) incapable of being lofted up to the region above the surface snowline because they have Stokes numbers $\St > \alpha$ \citep{youdin2007,ciesla2010}. This creates an imbalance between downward diffusion of gas-phase CO and upward mixing of CO-ice-rich grains, resulting in a removal of CO vapor in the warm upper parts of the disk on a timescale comparable to the vertical mixing timescale. In Fig.~\ref{fig:2D_gas}, we see that this depletion grows over time, ultimately reaching about ${\sim}90\%$ between radii of 50 and 100 au (see also Fig.~\ref{fig:snapshots1}(a)): at smaller radii, pebbles are still capable of (sometimes) reaching the warm upper parts, while outside of $r=100\mathrm{~au}$ pebble formation and vertical mixing are relatively slow compared to a million years (Fig.~\ref{fig:pbf}(c)).

This picture is somewhat analogous to the models of \citet{krijt2016b}, in which fragmentation-limited dust coagulation just outside the water snowline was found to lead to depletions of water vapor in the upper disk of up to a factor 50, with the (steady-state) magnitude of depletion increasing for decreasing $\alpha$. However, these strong depletions were only reached after dust coagulation had proceeded to lock most of the solid mass up in $1{-}10\mathrm{~cm}$-size particles \emph{and} sufficient time had passed to allow for vertical mixing. Because collisional fragmentation does not play a major role in the outer disk \citep[e.g., Fig.~\ref{fig:pbf}(b) and ][]{stammler2017}, it is not included in our simulations and no such steady-state is reached in model M0b and the fact that the depletion is smaller is merely a consequence of both the coagulation and mixing timescales being longer in the outer disk.

Because pebbles are not allowed to drift in model M0b, an effect similar to the one described above is operating in the radial direction, reducing the CO abundance just interior to the midplane snowline by several 10s of \% (see Figs.~\ref{fig:2D_gas} and \ref{fig:snapshots1}). This situation (outward diffusion of volatile followed by freeze-out and sequestration) is similar to the cold-finger effect discussed by \citet{stevenson1988}, as well as regime 3 in \citet[][Fig.~3]{cuzzi2004}. Lastly, the enhancement in the cold CO vapor visible just below the 22~K contour around $r=150\mathrm{~au}$ is larger compared to model M0a because the removal is small grains results in a longer freeze-out timescale, $\tau_\mathrm{fo} \propto \rho_\mathrm{d}^{-1}$, allowing more CO molecules to travel deeper into the disk before they are removed from the gas phase.

\subsection{Drifting pebbles}\label{sec:driftingpebbles}
\begin{figure}[t]
\centering
\includegraphics[clip=,width=1.\linewidth]{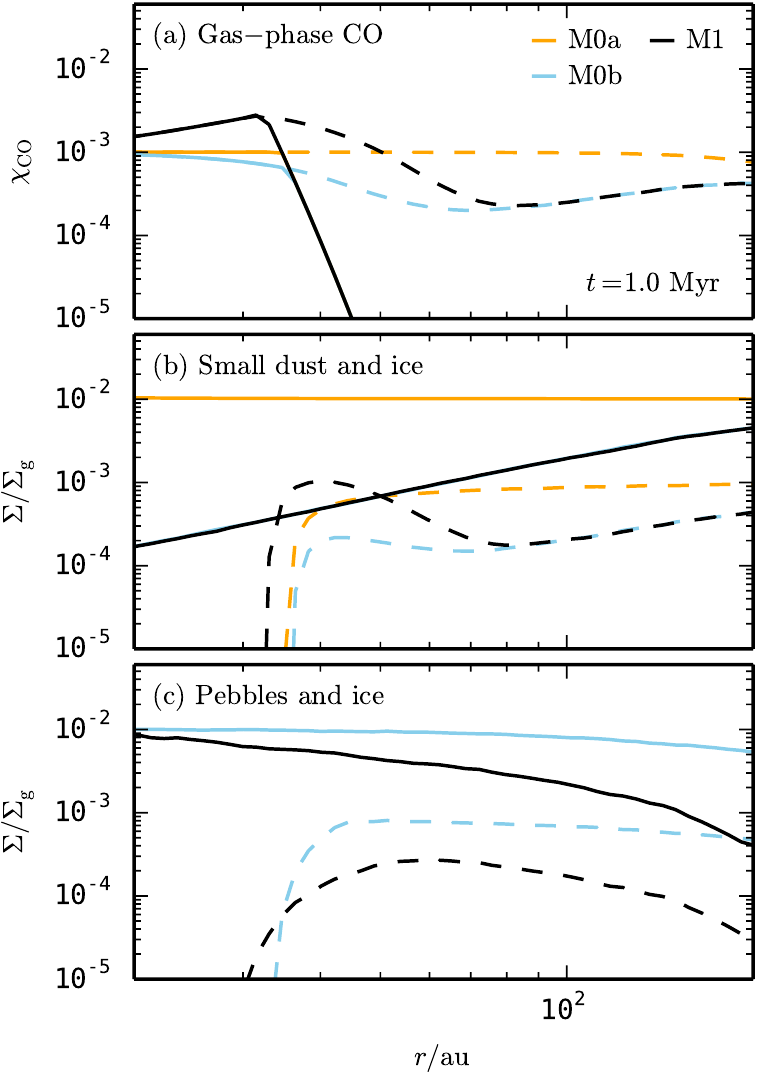}
\caption{Comparison between models M0a, M0b, and M1 at $t=1\mathrm{~Myr}$: (a) Gas-phase CO abundance in the midplane (solid) and warm (${>}22\mathrm{~K}$) region of the disk (dashed). (b) Surface densities of dust (solid) and CO-ice-on-small-dust (dashed). (c) Surface densities of pebbles (solid) and CO-ice-on-pebbles (dashed).}
\label{fig:snapshots1}
\end{figure}

In the final and most complete model of Sect.~\ref{sec:building} (model M1, which will serve as our standard model), we allow the formed pebbles to move vertically and drift radially according to Eq.~\ref{eq:r_t}. Focusing first on Fig.~\ref{fig:2D_sol}, we see that the inclusion of radial drift results in a decrease in the pebble abundance in the midplane (this is most clearly seen outside of $r>50\mathrm{~au}$) . The distribution of small dust above $z/H\sim1$ is similar to the one in model M0b however, because the timescale for converting dust into pebbles is the same. After 1~Myr, drift has reduced the pebble surface density by several 10s of \% at 50 au, and over 90\% at 200 au relative to model M0b (Fig.~\ref{fig:snapshots1}(c)). The magnitude of this reduction in the solid surface density is comparable to that obtained in 1D radial models including grain growth and radial drift \citep[e.g.,][]{birnstiel2012,lambrechts2014,stammler2017}.

As pebbles drift, they take the majority of the CO ice with them, generally reducing the ice abundance in the midplane (Fig.~\ref{fig:2D_ice}). When pebbles eventually drift through the CO snowline, they will sublimate and lose their CO ice. For large pebbles that drift rapidly, the timescale for ice-loss can become comparable to the radial drift timescale, resulting in them possibly traversing significant distances inside the snowline before losing all their ice \citep[e.g.,][]{piso2015,powell2017}. This results in (a fraction of) the ice on pebbles surviving interior to where ice was stable in model M0b (see Fig.~\ref{fig:snapshots1}(c)). Ultimately, the pebbles will lose all their CO ice, and their collective sublimation results in a plume of vapor that will spread vertically and radially. After 1~Myr, the midplane CO abundance interior to the snowline is increased by a factor of ${\sim}3$ (see Fig.~\ref{fig:snapshots1}), comparable to what was found by \citet[][Fig.~7]{stammler2017} for the case of $\alpha=10^{-3}$. As the CO abundance is increased locally, the location of the snowline moves inward by several au \citep[see also][]{stammler2017,powell2017}; this is seen most clearly in the distribution of ice on small grains in Fig.~\ref{fig:snapshots1}(b). As the plume of CO vapor spreads radially, part of it is mixed outward along the midplane, crossing the CO snowline in the opposite direction. This `retro-diffused' material preferentially freezes out on small grains \citep[see also][Fig.~3]{stammler2017}, resulting in a bump in the ice-on-small-dust distribution in Fig.~\ref{fig:snapshots1}(b). In fact, unlike in model M0b, the small grains between ${\sim}35-50\mathrm{~au}$ have an ice/rock ratio ${>}1$, and dominate the ice budget in the midplane (bottom row of Fig.~\ref{fig:2D_ice}), even if the solid mass is dominated by pebbles (bottom row of Fig.~\ref{fig:2D_sol}). At these locations, the CO ice mantles of small grains are almost 10x more massive in the simulation with drift compared to the simulation without drift (M0b).

A key advantage of our 2D model is that we can study how the plume of CO vapor expands into the upper regions of the disk. From the middle three rows of Fig.~\ref{fig:2D_gas} it is clear that while (most of) the CO molecules are released close to the midplane, vertical mixing works relatively quickly (from Fig.~\ref{fig:pbf}(c), $\tau_z\approx10^4\mathrm{~yr}$ at $20\mathrm{~au}$ for $\alpha=10^{-3}$) to smear out any vertical gradients in the CO abundance. In the radial direction, the region that shows an enhanced CO abundance (shown as blue in Fig.~\ref{fig:2D_gas}) grows steadily, extending far beyond where the midplane snowline is located. Focusing on the warm gas component after 1~Myr (Fig.~\ref{fig:snapshots1}(a)), the outward diffusion has a clear signature in the CO abundance of the warm disk component that can be seen out to ${\sim}80\mathrm{~au}$ (compared to model M0b), and results in a CO adundance that is elevated compared to the initial conditions as far out as $r=50\mathrm{~au}$.

\subsection{Pebble sizes}
\begin{figure}[t]
\centering
\includegraphics[clip=,width=1.\linewidth]{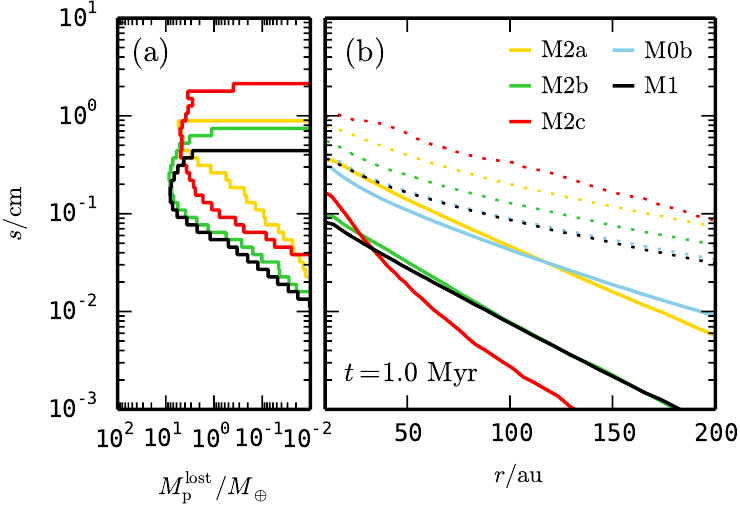}
\caption{(a) Integrated mass of pebbles of different sizes that have drifted interior to 10 au. (b) Radial profile of the mass-dominating size (solid) and maximum pebble size (dotted) at $t=1\mathrm{~Myr}$.}
\label{fig:pebbles_1}
\end{figure}

We can compare the distributions of physical pebble sizes in models M0b and M1. Figure~\ref{fig:pebbles_1}(b) shows the final radial profile of the mass-dominating\footnote{When calculating the mass-dominating size, the small dust is included and taken to have a size $s_\bullet$. In reality, these fractal grains will have a variety of sizes, most of which will be ${\gg}s_\bullet$. However, since the fractal grains behave as monomers, we deemed it appropriate to treat them as such for the purpose of this plot.} particle size (solid line) as well as the maximum pebble size (dotted). Comparing models M0b and M1, we see that while the maximum size is similar, the preferential removal of large grains by radial drift in model M1 has decreased the mass-dominating size by about an order of magnitude. Mass dominating sizes of the order of a millimeter/centimeter are a common outcome of dust evolution models \citep[e.g.,][]{birnstiel2012}.

In Fig.~\ref{fig:pebbles_1}(a), we show the cumulative mass of pebbles of different sizes that have exited the grid at the inner boundary at $r=10\mathrm{~au}$ over the course of the 1~Myr simulation. In model M1, a total of $65 M_\oplus$ of pebbles reached the inner disk, most of which had sizes between $0.5-5\mathrm{~mm}$. For comparison, at $t=0$, there are ${\approx}150M_\oplus$ of solids (excluding CO ice) present outside of $r=10\mathrm{~au}$ for our choice of disk mass and size. Both the total mass and the sizes of particles that reach the planet-formation region are important quantities in the context of planet formation through pebble accretion, because they set both the efficiency with which proto-planets can accrete material \citep[e.g.,][]{johansen2017}, as well as the total mass that is available for accretion. We discuss the variation in the properties of pebbles that reach the inner disk and the implications for pebble accretion models further in Sect.~\ref{sec:pebhist}.

\begin{figure*}[]
\centering
\includegraphics[clip=,width=1\linewidth]{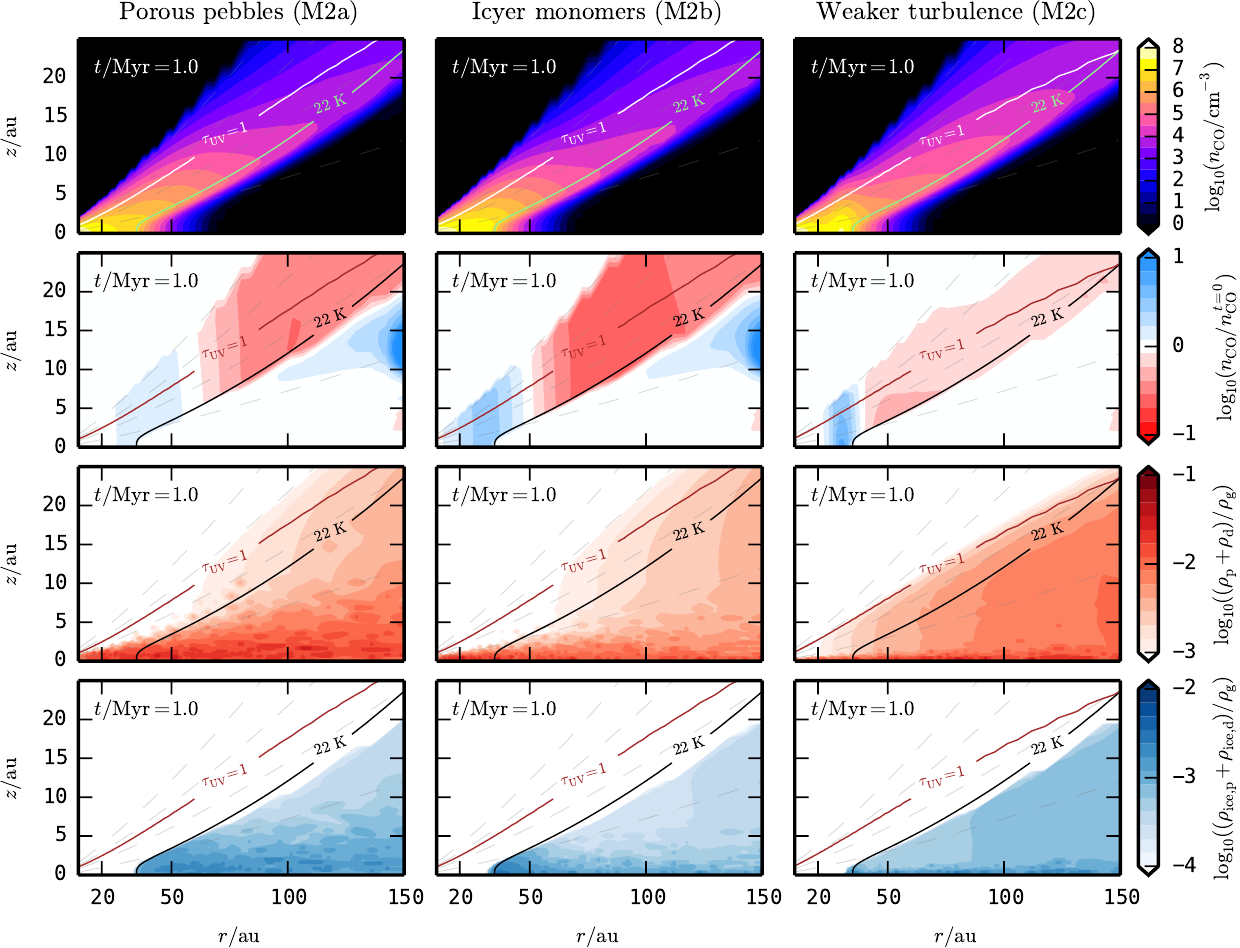}
\caption{Final distributions of gas-phase CO (top row), change in gas-phase CO (second row), solids (third row), and CO ice (fourth row) for models M2a, M2b, and M2c (see Table~\ref{tab:modelruns} and Sect.~\ref{sec:results}).}
\label{fig:2D_mix}
\end{figure*}

\section{Model sensitivity}\label{sec:results}
In this Section, we use model M1 as the basis for a small parameter exploration. In particular, we are interested in seeing how the turbulence strength and the assumptions that go into the dust evolution model influence the outcome of the calculations. To illustrate how changing various parameters impacts the observed behavior, we run a series of calculations where we vary key parameters one by one (see Table~\ref{tab:modelruns}). The results are plotted in Figs.~\ref{fig:2D_mix} and \ref{fig:snapshots2} and are described below. \new{While we focus here on studying the effects of changing the behavior of the dust and pebbles, the impact of varying the star+disk properties (e.g., stellar mass, disk size, mass, and temperature/density structure) on the CO redistribution will be the subject of a future study.}

\subsection{Pebble porosity}
In the context of our dust evolution model (Section \ref{sec:localpebbleformation}), the degree of pebble compaction in bouncing collisions (set by $\phi_\mathrm{c}$) plays a big role in determining the sizes and  Stokes number of the pebbles that are created locally: for a fixed particle mass, particle size scales as $s\propto\phi_\mathrm{c}^{-1/3}$ but the Stokes number as $\St \propto \phi_\mathrm{c}^{2/3}$. Decreasing the compactness by an order of magnitude (model M2a) will decrease the Stokes numbers of the formed pebbles by a factor ${\sim}5$. Consequences of the pebbles' Stokes numbers being smaller are: (i) vertical settling is less dramatic, making the population of solids and ices near the midplane more extended in the vertical direction (bottom two rows of Fig.~\ref{fig:2D_mix}). This also makes the sequestration of CO in the midplane more difficult and as a result the depletion of warm CO is less severe between $r=35{-}100\mathrm{~au}$ (Fig. \ref{fig:snapshots2}(a)); (ii) because radial drift is slower, pebbles remain abundant (Fig.~\ref{fig:snapshots2}(c)) and the flux of ices through the snowline is decreased, reducing the size of the plume of gas-phase CO that forms interior to the snowline (first two rows of Fig.~\ref{fig:2D_mix} and Fig.~\ref{fig:snapshots2}(a)).

\subsection{Monomer surface stickiness}
The extent to which water ice dominates the monomer surface ($\fw$) will also influence the Stokes numbers of pebbles that form locally: a higher $\fw$ increases the rolling energy, extending the duration of the fractal growth phase (Eq.~\ref{eq:m_roll}) and shifting the bouncing threshold velocity to higher aggregate masses (Eq.~\ref{eq:v_thr}), generally increasing the final pebble's Stokes number. In addition, there is a weak dependence of the pebble formation timescale $\tau_I$ through $m_\mathrm{roll}$, which is larger for water-ice covered grains (Sects. \ref{sec:stickiness} and \ref{sec:pebbleconversion}). Comparing models M2b ($\fw=1$) and M1 ($\fw=0.5$) in Fig.~\ref{fig:snapshots2}, we see that indeed the signatures of rapid radial drift (a decrease in the pebble surface density and a plume of CO vapor inside the snowline) become more evident for increasing $\fw$, but the differences are small.

\subsection{Turbulence strength}\label{sec:alpha}
\begin{figure}[t]
\centering
\includegraphics[clip=,width=1.\linewidth]{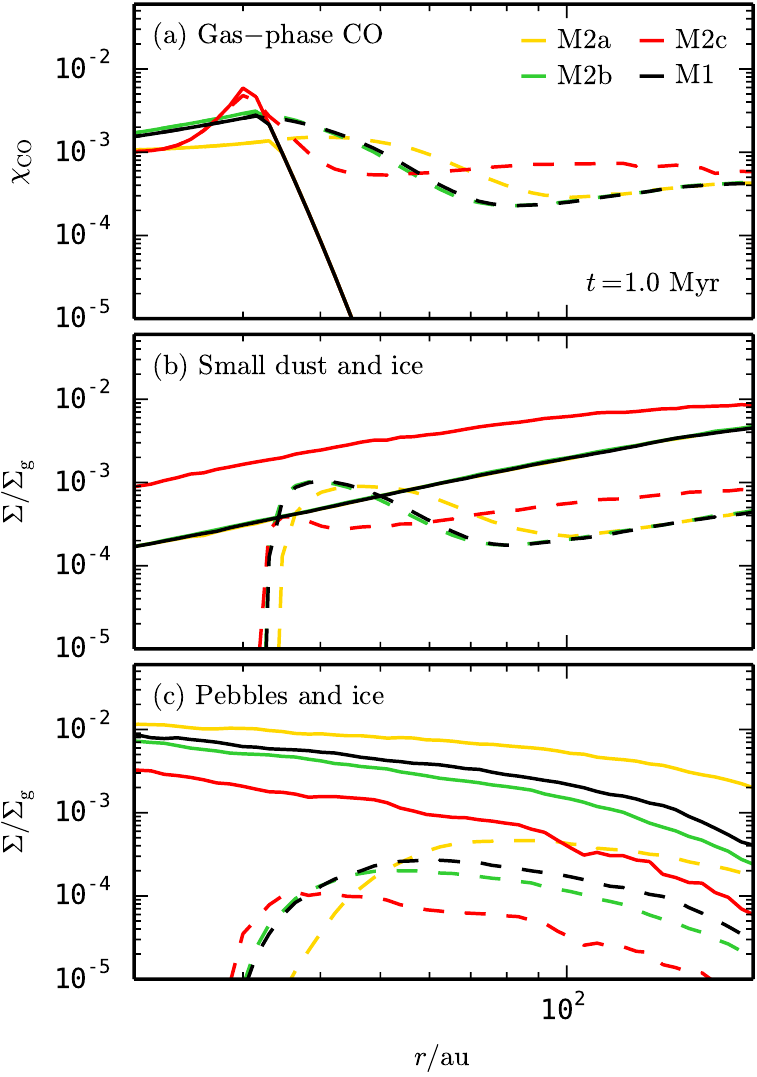}
\caption{Similar to Fig.~\ref{fig:snapshots1} but for models M2a through M2c (see Table \ref{tab:modelruns}). Model M1 is shown for comparison.}
\label{fig:snapshots2}
\end{figure}

Lastly, we vary the turbulence strength by lowering the value of $\alpha$ from $10^{-3}$ to $10^{-4}$ (model M2c). Having a weaker turbulence affects all aspects of the CO depletion/enhancement story. First, even without any grain growth, the importance of settling for the smallest grains increases \citep[e.g.,][]{dullemond2004} resulting in small grains already becoming depleted from the regions above $z/H\sim1.5$ (third row of Fig.~\ref{fig:2D_mix}). As small grains decouple from the gas around the surface snowline, the amount of CO depletion should increase (see Sect.~\ref{sec:M0a} and \citealt{xu2016}). The reason we do not see a more severe depletion at large radii in Fig.~\ref{fig:snapshots2}(a) is because decreasing $\alpha$ has increased the timescales involved: in our model, the diffusion coefficient is proportional to $\alpha$ (Sect.~\ref{sec:diskmodel}), which means that lowering $\alpha$ increases the timescales for vertical and radial diffusion of CO vapor and small dust grains. Specifically, the vertical mixing timescale $\tau_z \sim (\alpha \Omega)^{-1}$, so that at a radius of 100 au, $\tau_z\approx10^5\mathrm{~yr}$ for $\alpha=10^{-3}$, but $\tau_z\approx10^6\mathrm{~yr}$ for $\alpha=10^{-4}$. 

A weaker turbulence also affects the pebble formation and evolution process. First, due to the decrease in particle-particle collision velocities, the first two stages of the dust coagulation process (seen in Fig.~\ref{fig:dust_sketch}) are extended, leading to larger and more porous aggregates at the end of stage II. When these aggregates are compacted during stage III, the compact pebbles that are produced are larger (see Fig.~\ref{fig:pebbles_1}) -- and have higher Stokes numbers -- compared to those in Model M1 (the combined effects of pebbles being larger and turbulent mixing being weaker make the pebble sub-disk very geometrically thin and hard to see in Fig.~\ref{fig:2D_mix}). However, the timescale on which dust is converted into pebbles also becomes longer (see Sect. \ref{sec:pebbleconversion}). 

These effects together can explain the behavior observed in Figs.~\ref{fig:2D_mix} and \ref{fig:snapshots2}: the increase in pebble formation time leads to a high small-dust abundance and relatively low pebble surface density after 1~Myr, while the increase in the radial drift velocity of pebbles together with the smaller diffusion coefficient results in a narrower, higher peak for the gas-phase CO abundance in the midplane. For the CO vapor in the warm layer outside the midplane snowline, the weaker retro-diffusion leads to a relatively small CO abundance between $40{-}60\mathrm{~au}$, and while the amount of depletion in the outer disk would increase on long timescales, 1~Myr is too short for this depletion to occur, resulting in a higher CO abundance in the warm gas at large radii \citep[see also][Sect. 3.2]{xu2016}. Lastly, while the CO freeze-out timescale does not directly depend on $\alpha$, the weaker turbulence makes the freeze-out time shorter compared to the mixing timescale, effectively removing the enhancement in cold CO vapor below the surface snowline (second row of Fig.~\ref{fig:2D_mix}).

\begin{figure}[t]
\centering
\includegraphics[clip=,width=1.\linewidth]{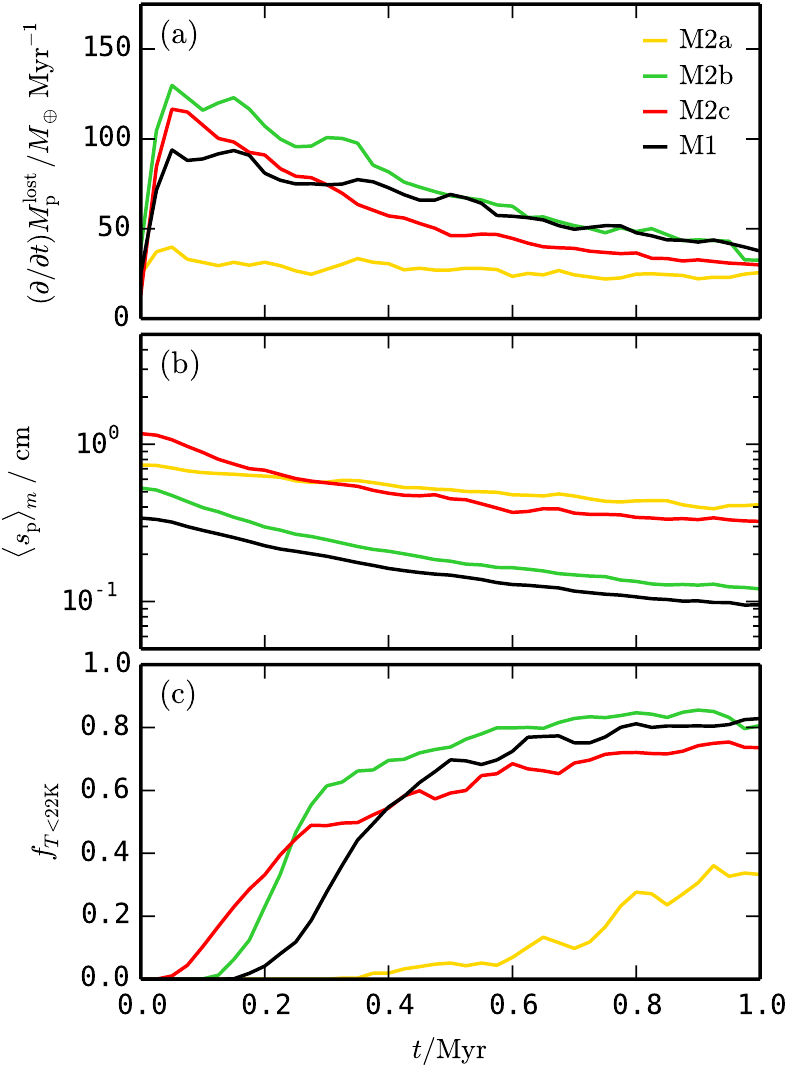}
\caption{Time evolution of properties of pebbles drifting through the inner boundary of our model grid $r=10\mathrm{~au}$: (a) total pebble mass flux; (b) mass-averaged size; (c) fraction of pebbles that formed in regions where $T<22\mathrm{~K}$.}
\label{fig:pebhistplot}
\end{figure}

\section{Pebbles reaching the inner disk}\label{sec:pebhist}
Apart from looking at the situation after 1~Myr of evolution (e.g., Fig.~\ref{fig:pebbles_1}), we can study how the amount and properties of pebbles that reach the inner disk change over time. These quantities are important in the context of planet formation through pebble accretion because the efficiency of the pebble accretion process depends sensitively on the aerodynamical properties (and hence the sizes) of the pebbles that are being supplied from the outer disk \citep[e.g.,][]{lambrechts2014,visser2016, johansen2017,ormel2018}.

Figure \ref{fig:pebhistplot}(a) shows the evolution of the (total) pebble mass flux through $r=10\mathrm{~au}$. The pebble flux, expressed in $M_\oplus/\mathrm{Myr}$, can be seen to vary significantly between models. In general, scenarios in which the Stokes numbers of formed pebbles are larger (i.e., high $\fw$, low $\alpha$) have the highest maximum pebble flux, which is achieved after ${\sim}10^5\mathrm{~yr}$. The flux then generally decreases with time. Qualitatively, this behavior is very similar to that seen in Fig.~7 of \citet{birnstiel2012}, although the fluxes we find are generally lower and do not decrease as rapidly. These differences are likely due to the fact \citet{birnstiel2012} used a more massive, smaller disk (with $M_\mathrm{disk}=0.1M_\odot$ and $r_c=60\mathrm{~au}$) as well as differences in the underlying dust coagulation model \new{(specifically, the treatment of aggregate porosity)}.

Figure \ref{fig:pebhistplot}(b) shows the mass-dominating size of pebbles reaching the inner disk as a function of time. Typical pebble sizes are between $0.1{-}1\mathrm{~cm}$, with the largest pebbles being supplied early on in simulations with weak turbulence (M2c) or sticky monomers (M2b). The model with decreased pebble compaction (M2a) also produces large grains, but in terms of their Stokes numbers these pebbles are much `smaller' because their internal density is reduced by 90\%. Over the course of 1~Myr, the typical size of pebbles crossing $r=10\mathrm{~au}$ drops by a factor ${\sim}2{-}5$, similar to the behavior observed in \citet[][Fig.~2]{lambrechts2014}. The models that exhibit the largest drop in particle size correspond to those that also show a large decrease in the pebble mass flux (Fig.~\ref{fig:pebhistplot}(a)).

Lastly, Figure \ref{fig:pebhistplot}(c) shows, again as a function of time, the mass fraction of pebbles arriving at $r=10\mathrm{~au}$ that have formed at temperatures below $22\mathrm{~K}$, i.e., outside the CO snowline. This plot illustrates the capability of the pebble tracer particle approach to follow where material originated from. In this particular example, we see that for model M2a almost no material from outside the CO snowline makes it to the inner disk in the first 0.5~Myr, while for models in which pebbles are born with larger Stokes numbers (e.g., M2b, or M2c), solids originating from outside the CO snowline dominate the mass of arriving material at times $t>0.2\mathrm{~Myr}$. Potential implications of these findings are discussed in Sect.~\ref{sec:discussion}.

\section{Discussion}\label{sec:discussion}

\subsection{Comparison to (resolved) CO observations}\label{sec:observations}
Apparent depletions of gas-phase CO in the outer disk have been reported by several authors for a variety of disks \citep[][]{favre2013, du2015, kama2016, mcclure2016, schwarz2016}, with depletion factors ranging from a factor of a few to 2 orders of magnitude. In addition, assuming a Solar (i.e., non-depleted) value for the $\co/\hyy$ mixing ratio results in (very) low gas disk masses and unusually high dust-to-gas ratios \citep{ansdell2016,eisner2016,miotello2016,miotello2017}. The depletions we observe in the warmer parts of the outer disk (e.g., Fig.~\ref{fig:2D_gas}) are typically around 90\%, but we discuss possibilities for creating more extreme depletion factors below.

Another result of our models that include drift is the formation of a plume of gas-phase CO interior to and around the midplane snowline. At least for TW Hya, such an obvious resurgence of CO is not seen by \citep{schwarz2016}. Even though there is a hint of an increase inside the snowline \citep[][Fig.~3(d)]{schwarz2016}, and the contrast between the CO abundance interior and exterior to the snowline is similar to what we predict (e.g., Fig.~\ref{fig:snapshots2}), \citeauthor{schwarz2016} find CO to be depleted on \emph{both} sides of the midplane snowline. While the tracer used by \citet{schwarz2016}, C$^{18}$O, is possibly optically thick in the inner disk, this picture of a lack of CO returning to the gas phase was confirmed by \citet{zhang2017} using the optically thin $^{13}$C$^{18}$O.

\subsection{Increasing the amount of CO depletion}\label{sec:moredepletion}
The models shown in this paper do not show CO depletions of more than an order of magnitude in the disk's surface layers. Here, we discuss \new{effects} that could potentially increase the depletion to reach the 2 orders of magnitude that have been reported for some disks.

\emph{Evolution over longer timescales.} The snapshots shown in Figs.~\ref{fig:2D_gas} and \ref{fig:2D_mix} do not represent a steady state: pebbles are continuously forming and the degree of CO depletion in the outer disk is increasing with time. Because pebbles are continuing to form and the vertical mixing timescale in the outer disk is ${\sim}10^{5}\mathrm{~yr}$ or longer (Sect.~\ref{sec:alpha}), running the models for a longer period of time is expected to increase the depletion.

\new{\emph{A vertical turbulence profile.} In this study, we have assumed a single, constant $\alpha$-value when describing the turbulent viscosity and diffusion coefficients (Sect.~\ref{sec:transport}). In reality, the strength and nature of the turbulence is expected to vary significantly between different regions in the disk \citep[e.g.,][]{turner2014}. Recent theoretical models studying the outer regions of protoplanetary disks tend to find a relatively weak turbulence in the midplane (corresponding to $\alpha \lesssim 10^{-3} $) and a stronger turbulence ($\alpha\sim10^{-2}$) in the upper layers \citep[e.g.,][and references therein]{simon2013a,simon2013b,bai2016}. The presence of such a vertical profile can significantly influence vertical transport of dust grains \citep{ciesla2010,ormel2018}, promoting the sequestration of icy bodies in the midplane and increasing the efficiency with which CO is removed from the gas-phase in the disk's upper regions \citep{xu2016}. Recent observational work, however, appears to show the turbulence in the upper layers of the disks around TW Hya and HD163296 is relatively weak \citep{teague2016,flaherty2017,flaherty2018}, implying $\alpha \sim 10^{-2}$ is not common in the surface layers of protoplanetary disks.}


\emph{Dust-pebble interactions.} Our dust evolution model does not include pebble mass gain/loss through collisions with much smaller particles. If the accretion of small grains is efficient however, this sweep-up could contribute to the depletion of dust and volatiles from the warm molecular layer: In the models shown in this paper, the only way for a CO molecule to end up on a pebble in the midplane is to freeze out on a small grain which then grows into a (previously non-existing) pebble. If sweep-up is efficient, a second route becomes available, in which a molecule freezes out onto a small grain which is subsequently accreted by an already-existing pebble. In regions of the disk where this second route is more efficient than the first (i.e., regions with a low dust density and/or high pebble surface density), the volatile depletion could then be much more dramatic. However, collisions between pebbles or aggregates and small dust grains do not necessarily result in sticking but can also lead to mass loss in the form of erosion or cratering \citep{schrapler2011,seizinger2013c,krijt2015}. If erosion is efficient, it might not only limit further growth of pebbles, but also be the dominant source of small grains at later times \citep{schrapler2018}, potentially alleviating the problems dust coagulation models often have in producing enough small grains to match multi-wavelength observations \citep[e.g.,][]{dullemonddominik2005,pohl2017}.

\emph{Chemistry.} Finally, we discuss the possibility of removing CO from the gas-phase by locally reprocessing CO through chemical reactions that lock the carbon in other molecules/species \citep[e.g.,][]{bergin2014,reboussin2015,yu2016,eistrup2017}. A recent comprehensive modeling study by \citet{schwarz2018} found that -- unless the cosmic ray rate is high -- it is difficult to deplete CO by an order of magnitude or more on a timescale of a million years, concluding that chemistry alone is not responsible for the majority of the observed depletions. Nonetheless, \new{several} models conducted at 100 au converted a significant fraction of CO to $\coo$-ice and $\mathrm{CH_3OH}$-ice on timescales shorter than a million years \citep[][Fig.~5]{schwarz2018}. With both mechanisms (chemical processing of CO and pebble-formation-mediated sequestration in the midplane) leading to an order of magnitude of CO depletion when acting on their own, it is tempting to imagine they can reach the observed two orders of magnitude when working together. In addition, while we focused exclusively on how dust growth impacted material transport, the coagulation of small grains into larger solids is also expected to alter the temperature profile and radiation field in the disk \citep[e.g.,][]{cleeves2016,facchini2017}. Developing models to understand how these physical and chemical processes interact will be the focus of future work.

\subsection{Decreasing the pile-up of CO interior to the snowline}\label{sec:lessreturn}
Observations do not appear to show a return of CO interior to the snowline \citep{schwarz2016}, the presence of which is a common outcome in our models that include both pebble formation and radial drift (Figs.~\ref{fig:2D_mix}). We briefly discuss possibilities that could prevent the CO from returning to the gas as pebbles grow and evolve.

\emph{Reduced drift efficiency.} Unsurprisingly, the models that show the smallest CO enhancement in the inner disk are those for which the radial flux of solids is smallest (cf. Figs.~\ref{fig:snapshots2}(a) and \ref{fig:pebhistplot}(a)). One way to reduce the pebble flux is to have the pebbles keep relatively small Stokes numbers, which, in the context of our dust evolution model, happens when pebbles maintain a high porosity (model M2a). Pebble sizes and Stokes numbers could also be kept small if catastrophic fragmentation is a common outcome of pebble-pebble collisions in the outer disk \citep{brauer2007,birnstiel2012,pinilla2017}, \new{as would be the case for $v_f\sim\mathrm{1~m/s}$ (Fig.~\ref{fig:pbf}(b))} . Alternatively, the efficiency of radial drift can be reduced by structures in the gaseous disk such as pressure bumps or traps \citep{pinilla2012}, which cause the pressure gradient $\eta$ (see Eq.~\ref{eq:driftsettle}) to vary on relatively small radial scales.

\new{\emph{Increased turbulence in the midplane.} The shape of the CO enhancement depends on the strength of the turbulence (compare models M1 and M2c in Fig.~\ref{fig:snapshots2}(a). The peak is less prominent for a higher value of $\alpha$ because $i)$ diffusion is more efficient at smearing out the deposited CO vapor and $ii)$ the individual sizes and the total radial flux of pebbles tend to decrease for higher $\alpha$ (see Fig.~\ref{fig:pebhistplot}). \citet{stammler2017} find that for turbulence strengths $\alpha \sim 10^{-2}$, the enhancement relative to the initial conditions becomes insignificant, although it is not clear if such high levels of turbulence are present in the disk midplane at radii outside ${\sim}30\mathrm{~au}$ \citep{simon2013a,simon2013b}. Alternatively, a lower Schmidt number would also increase the diffusivity and lead to a smaller peak in the CO abundance just interior to the snowline \citep[][Fig.~8]{stammler2017}.}

\new{\emph{High mass accretion rate.} With the gas accreting radially, the plume of CO vapor forming just inside the snowline will advect inward at a velocity $v_r \sim 3 \nu_\mathrm{T}/2r$ and result in the enhancement of the entire inner disk on a timescale comparable to the local viscous time. For the disk model outlined in Sect.~\ref{sec:diskmodel} and \ref{sec:transport}, $\dot{M}\sim10^{-9}~M_\odot/\mathrm{yr}$ and $v_r\sim \mathrm{cm/s}$ around the CO snowline and this effect can be ignored on the timescales simulated in Sects.~\ref{sec:building} and \ref{sec:results}. In disks with a higher accretion rate however, $v_r$ can become significant, decreasing the degree of vapor enhancement in the inner disk and the efficiency of CO vapor retro-diffusing back across the snowline \citep{cuzzi2004}.}

\emph{Planetesimal formation.} The only model in which we observe a depletion of CO vapor \new{inside the CO snowline} is one without any pebble migration (model M0b in Fig.~\ref{fig:snapshots1}(a)), in which case the pebbles outside the midplane snowline effectively become a sink for CO ice. While such a model does not appear to be realistic, a similar picture could arise if a large fraction of the pebbles can be converted into (stationary) planetesimals on timescales comparable to the drift timescale \citep[i.e., regime 3 of][]{cuzzi2004}.

\emph{Chemistry.} The explanations offered above all rely on decreasing the radial flux of pebbles, thus decreasing the flux of CO ice. A steady influx of solids could still be \new{allowed}, however, if CO can be destroyed chemically. \citet{schwarz2016} studied the chemical destruction of CO in the inner disk (at 19 au), finding that removing CO on a Myr timescale is only feasible with high cosmic ray rate. Alternatively, CO could be reprocessed already in the outer disk, before freezing out on the grains in the form of hydrocarbons or $\coo$ for example (see last paragraph of Sect.~\ref{sec:moredepletion}). However, while this might alleviate the apparent problem of not seeing the return of CO around $r\approx30\mathrm{~au}$, putting the carbon in $\coo$ will only make a similar issue at the $\coo$ snowline more severe \citep[see][]{bosman2017}. 

Developing models that include pebble formation and drift (this paper), chemical reactions involving the dominant carbon carriers \citep{schwarz2016} as well as planetesimal formation, and comparing those models to spatially resolved observations of nearby young disks will be key to understanding how carbon is delivered to the (terrestrial) planet formation zone \citep{bergin2014}.

\section{Summary}\label{sec:concl}
We have developed a two-dimensional (radial+vertical) model that describes the transport and interaction of gas-phase CO, small fractal dust grains, and larger mm/cm-size pebbles in protoplanetary disks on Myr timescales. We use this model to study how the large-scale formation and radial migration of pebbles impacts the gas-phase CO distribution on both sides of the snowline, in the disk midplane, and in the warmer upper regions of the (outer) disk. Our main findings are that:

\begin{itemize}

\item{The gas-phase CO abundance is variable in both time and space when dust coagulation proceeds to form pebbles that experience significant settling and radial drift (e.g., Figs.~\ref{fig:2D_gas}--\ref{fig:2D_ice}).}

\item{The formation and vertical settling of pebbles results in a depletion of CO vapor from the warm layer above the surface snow line (Fig.~\ref{fig:2D_gas}). \new{The depletion increases with time and its magnitude depends} on the timescales for dust coagulation and vertical mixing, and on the sizes of the pebbles that form. In the models considered here, the CO abundance in the warm gas layer ($T>22\mathrm{K}$) drops to $20{-}50\%$ of its original value after 1 Myr, (Figs. \ref{fig:snapshots1}(a) and \ref{fig:snapshots2}(a)).}

\item{The radial drift of CO-ice rich pebbles through the midplane snowline results in a plume of CO vapor just interior to the snowline (Figs.~\ref{fig:2D_gas} and \ref{fig:snapshots2}(a)). The size and shape of the plume depends on the strength of turbulent diffusion and the efficiency of radial drift (i.e., the sizes of the pebbles). In our models the maximum CO abundance is raised by a factor $2{-}6$.}

\item{The outward diffusion of this plume results in a peak in the ice content of small grains outside the snowline (Fig.~\ref{fig:snapshots1}(b)) and an increase in the gas-phase CO in the upper parts of the disk that can extend to radii 10s of au beyond the location of the midplane snowline (Fig.~\ref{fig:snapshots1}(a)).}

\end{itemize}
If the large-scale, sustained radial drift of pebbles is indeed an important and ubiquitous trait shared by most planet-forming disks, then the features described above should be commonplace. The absence, in particular of the plume of CO vapor interior to the snowline, could then point to the conversion of CO into a more refractory species, or to the radial mass flux of pebbles being drastically reduced by, for example, disk inhomogeneities or early planetesimal formation.

\acknowledgments
\new{SK would like to thank Chris Ormel, Djoeke Schoonenberg, Sebastian Stammler, Til Birnstiel, Jake Simon and Mihkel Kama for comments and enlightening discussions. The authors are also grateful to the reviewer for comments that helped improve the manuscript.} Support for Program number HST-HF2-51394.001 was provided by NASA through a grant from the Space Telescope Science Institute, which is operated by the Association of Universities for Research in Astronomy, Incorporated, under NASA contract NAS5-26555.



\end{document}